\documentclass[journal, twoside]{IEEEtran}
\usepackage{graphicx,subfigure,cite,stfloats,listings}
\usepackage{amssymb,amsmath,amsthm}
\usepackage{algorithm,algorithmic}
\usepackage{psfrag}

\usepackage[caption=false]{subfig}
\usepackage{blkarray, bigstrut} 
\usepackage[normalem]{ulem}

\newcommand*{\vertbar}{\rule[+0.5ex]{0.5pt}{2.5ex}}

\usepackage{multirow, tabularx, rotating}

\usepackage{tikz}
\usepackage{tikz-qtree}
\usetikzlibrary{shapes}

\usetikzlibrary{trees}
\usetikzlibrary{
	arrows.meta, 
	shapes.geometric, 
	calc, 
	shadows,
	decorations.text,
	backgrounds
	}\colorlet{linecol}{black!75}

\usepackage{forest}
\usepackage{newfloat}
\DeclareFloatingEnvironment[fileext=lod]{Figure} 

\forestset{/.style=
{
	{for tree=
		{parent anchor=south, child anchor=north, align=center l+=1.5cm}
		}
	}
}

\pgfkeys{/forest,
	my rounded corners/.append style={rounded corners=2pt},
}

\usetikzlibrary{
	arrows, 
	decorations.markings, positioning}

\makeatletter

\usepackage{color, dsfont, bbm}
\usepackage[mathscr]{euscript}

\providecommand{\keywords}[1]{\textbf{\textit{Index terms---}}#1}

\DeclareMathAlphabet{\mathpzc}{OT1}{pzc}{m}{it}
\DeclareMathAlphabet{\mathitsf}{OML}{cmbr}{m}{it}

\newtheorem{example}{Example}
\newtheorem{remark}{Remark}

\DeclareMathOperator{\E}{\mathds{E}}

\newcommand{\B}[1]{\mathbf{#1}}

\newcommand{\EX}[1]{\E\left\{{#1}\right\}} 
\newcommand{\EXs}[2]{\E_{{#1}}\left\{{#2}\right\}} 

\DeclareFixedFont{\ttb}{T1}{txtt}{bx}{n}{12} 
\DeclareFixedFont{\ttm}{T1}{txtt}{m}{n}{12}  
\definecolor{deepblue}{rgb}{0,0,0.5}
\definecolor{deepred}{rgb}{0.6,0,0}
\definecolor{deepgreen}{rgb}{0,0.5,0}

\newcommand\pythonstyle{\lstset{
		language=Python,
		basicstyle=\ttm,
		otherkeywords={self},             
		keywordstyle=\ttb\color{deepblue},
		emph={MyClass,__init__},          
		emphstyle=\ttb\color{deepred},    
		stringstyle=\color{deepgreen},
		frame=tb,                         
		showstringspaces=false            %
}}

\lstnewenvironment{python}[1][]{
	\pythonstyle
	\lstset{#1}
}{}

\newcommand\pythoninline[1]{{\pythonstyle\lstinline!#1!}}

\usepackage{booktabs}
\usepackage{varwidth}

\makeatletter
\def\endthebibliography{%
	\def\@noitemerr{\@latex@warning{Empty `thebibliography' environment}}%
	\endlist
}

\usepackage{multirow}
\usepackage{xcolor}
\usepackage{colortbl}

\usepackage{slashbox}  

\usepackage{bigstrut}

\usepackage{supertabular}

\begin{document}
\title{
Physical Layer Authentication and Security Design in the Machine Learning Era}
\author{
	\IEEEauthorblockN{
		Tiep M. Hoang, 
		Alireza Vahid,
            Hoang Duong Tuan,
		and Lajos Hanzo
	}

	\thanks{T. M. Hoang and A. Vahid are with the Department of Electrical Engineering, University of Colorado Denver, Denver, CO 80204, USA (e-mails: minhtiep.hoang@ucdenver.edu; alireza.vahid@ucdenver.edu).}
	\thanks{H. D. Tuan is with the School of Electrical and Data Engineering, University of Technology Sydney, Broadway, NSW 2007, Australia (email:Tuan.Hoang@uts.edu.au). }
	\thanks{L. Hanzo is with the School of Electronics and Computer Science, University of Southampton, Southampton SO17 1BJ, U.K. (email: lh@soton.ac.uk).}
}
	
\maketitle
	
\begin{abstract}
Security at the physical layer (PHY) is a {salient} research topic in wireless systems, and machine learning (ML) {is emerging} as a powerful tool for providing new data-driven security solutions. Therefore, the application of ML techniques to the PHY security is of crucial importance in the landscape of more and more data-driven wireless services. In this context, we first summarize {the family of bespoke} ML algorithms that are {eminently suitable for} wireless security. Then, we review the recent {progress in} ML-aided PHY security, where the term ``PHY security'' is classified into two different types: i) PHY authentication and ii) secure PHY transmission. Moreover, we treat neural networks as special types of ML and present how to deal with PHY security optimization problems using neural networks. Finally, we identify some major challenges and opportunities {in tackling PHY security challenges by applying carefully tailored ML tools.}
\end{abstract}
	
		


{\small
\begin{description}[\IEEEsetlabelwidth{so so long}\IEEEusemathlabelsep]
\item[AE] {Auto-encoder}
\item[Compl-NN] Complex-valued neural network
\item[Compl-OPT] Complex-valued optimization
\item[CSI]  Channel state information
\item[DOA] Direction of arrival
\item[GAN] Generative adversarial network
\item[GD] Gradient descent
\item[IoT] Internet-of-Things
\item[$k$-means] $k$-means clustering
\item[$k$-NN] $k$-Nearest neighbour
\item[LDA] Linear discriminant analysis
\item[LED] Light-emitting diode
\item[LiFi] Light fidelity
\item[MIMO] Multiple-input multiple-output
\item[ML] Machine learning
\item[MLP] Multiple layer perceptron
\item[mmWave] Millimeter wave
\item[MOO] Multiple-objective optimization
\item[MUSIC] MUlti-SIgnal Classification
\item[NLoS] Non-line-of-sight
\item[OC-SVM] One-class support vector machine
\item[PGD] Projected gradient descent
\item[PHY] Physical layer
\item[RBF] Radial basis function
\item[Real-NN] Real-valued neural network
\item[Real-OPT] Real-valued optimization
\item[RIS] Reconfigurable intelligent surface
\item[RRN] Recurrent neural network
\item[RSS] Received Signal Strength
\item[SGD] Stochastic gradient descent
\item[SSL] Self-supervised learning
\item[SVM] Support vector machine
\item[TOA] Time of arrival
\item[VLC] Visible light communication
\end{description}
}
\section{Introduction}\label{SEC: INTRO}
\subsection{Physical Layer Security}
Physical layer (PHY) security has become a new line of research independent of the security at higher layers \cite{Zou2016, Trappe_Survey_Security_2015, Mukherjee_Survey_Security_2014}. The basic foundation of PHY security lies in the randomness of the propagation environment {and of hardware impairments,} which are hard to {mimic}. {In parallel to the evolution} of wireless communications, PHY security has been investigated in a wide variety of communication systems. {Numerous} PHY security methods have been developed, which can be  categorized into the following {pair of} areas:
\begin{itemize}
	\item \textbf{PHY authentication}: Upon receiving a signal, authentication is needed to {ascertain} whether it comes from a trusted source or not. Through authentication, the origin of a signal can be tracked and the presence of an illegal user can be identified. In PHY authentication, channel uniqueness between a pair of devices can be exploited as a means of {characterizing} their identities and positions \cite{Trappe_Survey_Security_2015}. A pair of common attacks in PHY authentication are jamming and spoofing attacks. The purpose of jamming attacks is to contaminate the legitimate transmission. In contrast, the purpose of spoofing is to deceive the receiver into believing they come from a {legitimate source}. These two types of signals are generated by active eavesdroppers, who seek to break into a communication system.  
	\item \textbf{PHY security design}: This refers to the way we deal with security vulnerabilities through designing secure transmission strategies as well as systems at the physical layer. From a secure transmission design perspective, PHY security designs {may} incorporate {quite} different techniques {to guard} against eavesdropping, such as {directional} beamforming techniques \cite{Security_Beamforming_2018}, artificial noise \cite{Security_AN_2017}, multiple antennas \cite{Mutiple-antenna_Security_Zou2015}, and transmit antenna selection \cite{Security_TAS_2018}. From a network design perspective, PHY security designs also refer to the inclusion, selection, and cooperation of network entities for enhancing the security, such as the cooperation of relay nodes to form cooperative beamforming \cite{Cooperative-beamforming_Security_2019} or the employment of an intermediate node as a friendly jammer \cite{TMHoang-WCL-2017}. Optimizing the performance of a secure system is another aspect of the PHY security design \cite{Security_OPT_Survey2019}. The security can be optimized in a number of ways subject to a given set of {optimization} constraints. For example, a communication system can be designed {for ensuring} that the probability of information leakage is minimized within a fixed power budget or vice versa, and/or the power consumption is minimized while still guaranteeing the {target} security level.  
\end{itemize}

To {quantify} how well a secure system performs, most of the papers on PHY security {tend to rely on} a few performance metrics, such as the secrecy rate, secrecy outage probability, and secure energy efficiency \cite{Performance-metrics_Security_2016, Secure-Energy-Efficiency_Security2016}. The most typical metric is the secrecy rate {given by} the difference between the capacity of the legitimate channel and the capacity of the wiretap channel. It is worth mentioning that {the specific choice of the} performance metrics {plays a more influential role} in the context of PHY security design {than the particular choice of the optimization tool itself}. 

When it comes to eavesdropping, typically two types of eavesdroppers are considered: i) passive eavesdroppers who only listen {but never transmit}, and ii) active eavesdroppers who seek {to glean} confidential information {by applying malicious} tricks \cite{Passive-and-Active-Eve_Survey2016}. The detection of passive eavesdroppers seems to be {at first sight} impossible due to the fact that they do not transmit any signal. For this reason, passive eavesdroppers are {rarely} considered in PHY authentication. By contrast, both passive and active eavesdroppers are considered in PHY security design, because {the security-level of the} transmission strategy/policy will directly affect on the {likelihood} of {successfully extracting} confidential information by the eavesdroppers. 

\subsection{ML-based Physical Layer Security}
The development of ML has {spanned} several decades {and as a recent benefit, it has led to} successful applications in different fields. Hence, the introduction of ML into wireless networks {is at the time of writing} believed to lead to intelligent systems {having intelligent} learning capabilities \cite{Pareto_Wang2020}. As part of wireless communications, PHY security is also expected to {beneficially} exploit ML for further enhancing the security. Although the exploitation of ML in PHY security has been studied (e.g., in \cite{PHY-Authentication_Reinforcement, PHY-Authentication_ML_1, PLS_Information-theoretic_Besser2019}), {it is still in its infancy} compared to the extensive {body of} investigations of PHY security methods {dispensing with ML}. In fact, there are {numerous knowledge} gaps, which together form a new area of research, namely \emph{machine-learning-aided physical-layer security} (ML-aided PHY security). {As a dual pair of} the two main branches of PHY security, ML-aided PHY security also encompasses two branches of research: i) ML-aided PHY authentication, and ii) ML-aided PHY security design. 

As for ML-aided PHY authentication, the commonly-used tools of detecting malicious intrusions at the physical layer are supervised and unsupervised classification algorithms (inclusive of neural-network-based classification algorithms) \cite{ PHY-Authentication_ML_1, PHY_authentication_Wang2017, PHY_authentication_7_people, PHY-Authentication_ML_3, PHY-Authentication_ML_4, PHY-Authentication_ML_GaussianMixture_2018, PHY_authentication_Logistic_Regression, PHY-Authentication_KLT_2019, PHY-Authentication_Spoofing-attacks_2018  }. {Indeed,} many {of the ML-assisted techniques} do not even have to set up a fixed threshold for distinguishing {the related} data, while traditional PHY authentication {solutions critically} rely on a threshold and appear to be sensitive to the change in its value {upon making compromised/uncompromised decision} \cite{PHY_authentication_Wang2017, PHY-Authentication_ML_GaussianMixture_2018}. This {suggests that} the traditional methods, which typically rely on hypothesis testing, are {vulnerable to} non-stationary data. Thus, reinforcement learning--which is {a salient} branch of ML--{may} also be used for detecting eavesdropping attacks in dynamically {fluctuating} environments \cite{PHY-Authentication_Reinforcement}. 

As for ML-aided PHY security design, neural networks and reinforcement learning {constitute} the most popular approaches \cite{He2019-PLS-ML, PLS_Information-theoretic_Besser2019, PLA_Xing2019, PLS_Reinforcement_Li2019, ML-PHY-Design_Reinforcement_Xiao2019, PLS_Reinforcement_Miao2019}. Having said that, there {are some contributions} considering the potential use of other ML types, for example, support vector machine (SVM) and naive Bayesian {solutions} \cite{SVM_He2018}. Moreover, it is common to see optimization problems in the process of designing secure communication systems, {which lend themselves to} ML-aided security design \cite{He2019-PLS-ML, PLS_Information-theoretic_Besser2019, PLA_Xing2019, PLS_Reinforcement_Li2019, ML-PHY-Design_Reinforcement_Xiao2019, PLS_Reinforcement_Miao2019, SVM_He2018 }. 

In general, the aforementioned {treatises} on ML-aided PHY security help to shed light on the potential use of ML for securing communication systems from the physical-layer perspective. However, the overall ML-aided PHY security {of wireless systems} is still incomplete, especially compared to the mature conventional PHY security methods. {Hence, there is an urgent need to critically appraise all} to guide and {inspire future research}. 
\subsection{Previous Works}
There {is already substantial literature} on the application of ML in {general communication problems}, e.g., \cite{Pareto_Wang2020, Survey_Deep_Mao2018, Diro2018-RNN, 6805162, Cyber_ML_Buczak2016}; however, they do not focus on the topic of PHY security. 
By considering the future trends of wireless systems at different layers (including both the physical layer and the data link layer), the authors of \cite{Survey_Deep_Mao2018} focus {their attention merely} on deep learning {rooted in} neural networks. 
Diro and N. Chilamkurti \cite{Diro2018-RNN} {concentrated on} long short-term memory networks that originate from recurrent neural networks but the topic of \cite{Diro2018-RNN} is related to cyber security, {rather than on} PHY security. 
Alsheikh \textit{et al.} \cite{6805162} applied ML algorithms {for optimizing} wireless sensor networks. 
{While Buczak and Guven} \cite{Cyber_ML_Buczak2016} mainly {address} cyber security issues in wired networks. 
{However, the important class of} anomaly detection algorithms is not mentioned in  \cite{Pareto_Wang2020, Survey_Deep_Mao2018}. 
Only {the treatises} \cite{Diro2018-RNN}, \cite{6805162} and \cite{Cyber_ML_Buczak2016} {consider} anomaly detection, but {with no particular attention dedicated to} PHY security. 
{However, there is a paucity of literature on} ML-aided PHY security, {with the exception of} \cite{Survey_Deep_Chen2019, Survey_PLA_Fang2019}. 

\begin{table*}[t!]
\centering 
\caption{A comparison between this survey and other previous surveys.}
\renewcommand{\arraystretch}{1.75}
\begin{tabular}[1\textwidth]{ |c |c| *{10}{c|} c| }
\cline{3-13}
\multicolumn{2}{c|}{\multirow{2}{*}{ }}
&\multicolumn{10}{c|}{ Tutorial-and-survey papers } 
&
\\ \cline{3-12}
\multicolumn{2}{c|}{}
& \cite{Cyber_ML_Buczak2016}
& \cite{Survey_Fingerprints_Baldini2017} 
& \cite{Survey_Deep_Mao2018}  
& \cite{Survey_SmartGrid_Islam2019} 
& \cite{Survey_ML_Security_2020_rival_1}  
& \cite{1-Good-survey-2021} 
& \cite{2-Liu-Survey-new}  
& \cite{3-Mahdi-Survey-new}  
& \cite{4-Talpur-Survey-new} 
& \cite{ruzomberka2023challenges} 
& This
\\
\multicolumn{2}{c|}{}
& 2016 
& 2017
& 2018
& 2019
& 2020
& 2021
& 2022
& 2022
& 2022
& 2023
& survey
\\ \hline
\multicolumn{2}{|l|}{ \textbf{ML-aided PHY security} } 
& {} 
& $\checkmark$ 
& {} 
& $\checkmark$ 
& $\checkmark$ 
& $\checkmark$ 
& $\checkmark$ 
& {} 
& {} 
& $\checkmark$ 
& $\checkmark$ 
\\ \hline
\multicolumn{2}{|l|}{ \textbf{ML-aided cyber security} } 
& $\checkmark$ 
& {} 
& $\checkmark$ 
& {} 
& $\checkmark$ 
& {} 
& {} 
& $\checkmark$ 
& $\checkmark$ 
& {} 
& {} 
\\ \hline \hline  
\multicolumn{2}{|l|}{ \textbf{ML-aided PHY Authentication} }
& {} 
& $\checkmark$ 
& {} 
& $\checkmark$ 
& $\checkmark$ 
& $\checkmark$ 
& $\checkmark$ 
& {} 
& {} 
& {} 
& $\checkmark$ 
\\ \hline
\multicolumn{2}{|l|}{ \textbf{ML-aided PHY Secure Transm.} }
& {} 
& {} 
& {} 
& $\checkmark$ 
& {} 
& $\checkmark$ 
& {} 
& {} 
& {} 
& {} 
& $\checkmark$ 
\\ \hline \hline  
\multicolumn{2}{|c|}{ \multirow{2}{*}{ \shortstack{ \textbf{Neural Network-Aided} \\ \textbf{Security Optimization} } } } 
& \multirow{2}{*}{} 
& \multirow{2}{*}{} 
& \multirow{2}{*}{} 
& \multirow{2}{*}{} 
& \multirow{2}{*}{} 
& \multirow{2}{*}{} 
& \multirow{2}{*}{} 
& \multirow{2}{*}{} 
& \multirow{2}{*}{} 
& \multirow{2}{*}{} 
& \multirow{2}{*}{$\checkmark$}  
\\ 
\multicolumn{2}{|l|}{ } 
& {} 
& {} 
& {} 
& {} 
& {} 
& {} 
& {} 
& {} 
& {} 
& {} 
& {}  
\\ \hline \hline
\multirow{5}{*}{  \shortstack{ \textbf{PHY-related}\\ \textbf{features} } } 
& TOA  
& {} 
& {} 
& {} 
& {} 
& {} 
& {} 
& {} 
& {} 
& {} 
& {} 
& $\checkmark$ 
\\ \cline{2-12}
& DOA 
& {} 
& {} 
& {} 
& {} 
& {} 
& {} 
& {} 
& {} 
& {} 
& {} 
& $\checkmark$ 
\\ \cline{2-12}
& RSS 
& {} 
& {} 
& {} 
& {} 
& {} 
& {} 
& $\checkmark$ 
& {} 
& $\checkmark$ 
& {} 
& $\checkmark$ 
\\ \cline{2-12}
& Channel Gain 
& {} 
& {} 
& {} 
& {} 
& {} 
& {} 
& $\checkmark$ 
& {} 
& {} 
& {} 
& $\checkmark$ 
\\ \cline{2-13}
& Fingerprints 
& {} 
& $\checkmark$ 
& {} 
& {} 
& $\checkmark$ 
& {} 
& $\checkmark$ 
& {} 
& $\checkmark$ 
& {} 
& $\checkmark$ 
\\ \hline \hline 
\multirow{7}{*}{ 
\shortstack{
\textbf{Discussion} \\ \textbf{of ML} \\ \textbf{algorithms} 
} } 
& K-NN 
&Detailed 
&Detailed
& {} 
& {} 
&Detailed
& {} 
& Limited 
& Limited 
& Detailed 
& {} 
&Detailed
\\ \cline{2-13}
& SVM
&Detailed
&Detailed
& Limited 
& {} 
&Detailed
& {} 
& Limited 
& {} 
& Detailed 
& {} 
&Detailed
\\ \cline{2-13}
& LDA 
& {} 
& Limited 
& {} 
& {} 
& {} 
& {} 
& {} 
& {} 
& {} 
& {} 
&Detailed
\\ \cline{2-13}
& K-Means
& Limited 
& Limited 
& {} 
& Limited 
&Detailed
& {} 
& Limited 
& Limited 
& Limited 
& {} 
&Detailed
\\ \cline{2-13}
& One Class SVM 
&Detailed
& {} 
& {} 
& {} 
& {} 
& {} 
& {} 
& {} 
& {} 
& {} 
&Detailed
\\ \cline{2-13}
& Random Forest 
&Detailed
& {} 
& {} 
& {} 
&Detailed
& {} 
& Limited 
& {} 
& Detailed 
& {} 
&Detailed
\\ \cline{2-13}
& Isolation Forest 
& {} 
& {} 
& {} 
& {} 
& {} 
& {} 
& {} 
& {} 
& {} 
& {} 
&Detailed
\\ \cline{2-13}
& Hierarchical
& \multirow{2}{*}{Limited} 
& \multirow{2}{*}{Limited} 
& {} 
& {} 
& {} 
& {} 
& {} 
& {} 
& {} 
& {} 
& \multirow{2}{*}{Detailed} 
\\ 
& Clustering
& {} 
& {} 
& {} 
& {} 
& {} 
& {} 
& {} 
& {} 
& {} 
& {} 
&{} 
\\ \cline{2-13}
& Reinforcement 
& {} 
& {} 
& \multirow{2}{*}{Detailed} 
& {} 
& \multirow{2}{*}{Detailed} 
& {} 
& \multirow{2}{*}{Limited} 
& {} 
& \multirow{2}{*}{Detailed} 
& {} 
& \multirow{2}{*}{Limited} 
\\
& Learning 
& {} 
& {} 
& {} 
& {} 
& {} 
& {} 
& {} 
& {} 
& {} 
& {} 
& {} 
\\ \cline{2-13}
& Neural Networks 
&Detailed
&Detailed
&Detailed
& Limited 
&Detailed
& {} 
& Limited 
& Limited 
& Detailed 
& {} 
&Detailed
\\ \hline 
\end{tabular}
\label{tab: SOA comparison}
\end{table*}

In particular, {Chen} \textit{et al.} \cite{Survey_Deep_Chen2019} characterize the capability of deep learning to extract attack features in a cyber-physical transportation system. 
{By contrast, Fang} \text{et al.} \cite{Survey_PLA_Fang2019} portray some challenges to be faced in ML-aided PHY authentication. 
{However, these short magazine papers are constrained to an outline of the vast realm of} PHY security.
{A more detailed} PHY security {landscape was portrayed by} Baldini \textit{et al.} in \cite{Survey_Fingerprints_Baldini2017} and by Islam \textit{et al.} in \cite{Survey_SmartGrid_Islam2019}. 
To elaborate, \cite{Survey_Fingerprints_Baldini2017} focuses on {exploiting the unique} hardware fingerprints of physical devices as features and {appraises} the potential of some ML classification algorithms for PHY authentication. By contrast, Islam \textit{et al.} \cite{Survey_SmartGrid_Islam2019} surveys PHY security vulnerabilities and countermeasures {in the context of} smart energy systems. 

{A few years ago, Al-Garadi} \textit{et al.} \cite{Survey_ML_Security_2020_rival_1} surveyed security of the Internet of Things and occasionally mentioned PHY authentication, but {the prevalent} security issues at the physical layer are not addressed. 
Nguyen \textit{et al.} \cite{1-Good-survey-2021} outline the current challenges to be faced in the PHY security and cover a wide range of technical solutions. However, the ML techniques as well as ML-aided security solutions are not emphasized. Liu \textit{et al.} \cite{2-Liu-Survey-new} survey {the benefits of} ML algorithms in Internet-of-Things (IoT) device identification, {but they only focus on the issues of} PHY authentication, {while ignoring} secure PHY transmission between transceivers.
On the other hand, the pair of recent survey papers \cite{3-Mahdi-Survey-new, 4-Talpur-Survey-new} are inherently intended for ML-aided cyber-security, rather than ML-aided PHY security, although {the authors briefly} touch upon some of the PHY security risks.

Last but not least, PHY-security-related optimization problems are not presented in \cite{ Pareto_Wang2020, Survey_Deep_Mao2018, Diro2018-RNN, 6805162, Cyber_ML_Buczak2016, Survey_Deep_Chen2019, Survey_PLA_Fang2019, Survey_ML_Security_2020_rival_1, Survey_Fingerprints_Baldini2017, Survey_SmartGrid_Islam2019, 1-Good-survey-2021, 2-Liu-Survey-new, 3-Mahdi-Survey-new, 4-Talpur-Survey-new }. 
Given that neural networks can be exploited for efficiently solving stochastic optimization problems, it is promising to use neural networks for optimizing PHY security. This is a challenge to be {tackled, but this opportunity was missed} by the authors of \cite{ Pareto_Wang2020, Survey_Deep_Mao2018, Diro2018-RNN, 6805162, Cyber_ML_Buczak2016, Survey_Deep_Chen2019, Survey_PLA_Fang2019, Survey_ML_Security_2020_rival_1, Survey_Fingerprints_Baldini2017, Survey_SmartGrid_Islam2019, 1-Good-survey-2021, 2-Liu-Survey-new, 3-Mahdi-Survey-new, 4-Talpur-Survey-new }. By contrast, the subject of using neural networks for handling PHY security optimization {problems constitutes} an important part of this survey. 
Table \ref{tab: SOA comparison} highlights the major differences between this work and the related survey papers.





 
\subsection{Contributions}
When it comes to the {employment of} ML in PHY security, we consider the following {pair of pivotal} goals: 
\begin{itemize}
	\item \textbf{GOAL 1}: Harnessing \emph{ML classification} algorithms for PHY authentication and for PHY security designs. 
	\item \textbf{GOAL 2}: Invoking (artificial) \emph{neural networks} for PHY security with a focus on design optimization. 
\end{itemize}
Motivated by the fact that {there is a paucity of literature on} these challenging yet promising goals, {especially in terms of} survey and tutorial papers, we aim {for filling this gap.} Our main contributions are summarized as follows:
\begin{itemize}
	\item We {commence by reviewing the typical} primary data sources {routinely} employed in wireless communication systems for training ML algorithms.   
	\item {From the rich family of} ML classification algorithms, we review four typical supervised learning algorithms, four typical unsupervised learning algorithms, and neural networks that may belong either to the family of supervised or unsupervised learning. We {dedicate particular attention to} anomaly detection methods that {have been} ignored in previous surveys. Additionally, we discuss the application potential of ML classification in meeting \textbf{GOAL 1}.  
	\item Regarding \textbf{GOAL 2}, we bridge the gap between PHY security optimization and the ability of neural networks in dealing with stochastic optimization. Although the employment of neural networks for solving optimization problems {has found favour} in ML research, it has not been {popularized} in secure communications. Hence, this is the first survey {critically appraising} the application potential of neural networks in optimizing PHY security designs. Furthermore, we also reveal the ability of neural networks to optimize communication system designs. {In this vast field,} PHY security designs {constitute a particularly important instance, because they must operate in the face of uncertainty, where conventional techniques fail.}
\end{itemize}

\subsection{Organization}
\begin{figure*}[!t]	\centerline{\includegraphics[width=1\linewidth]{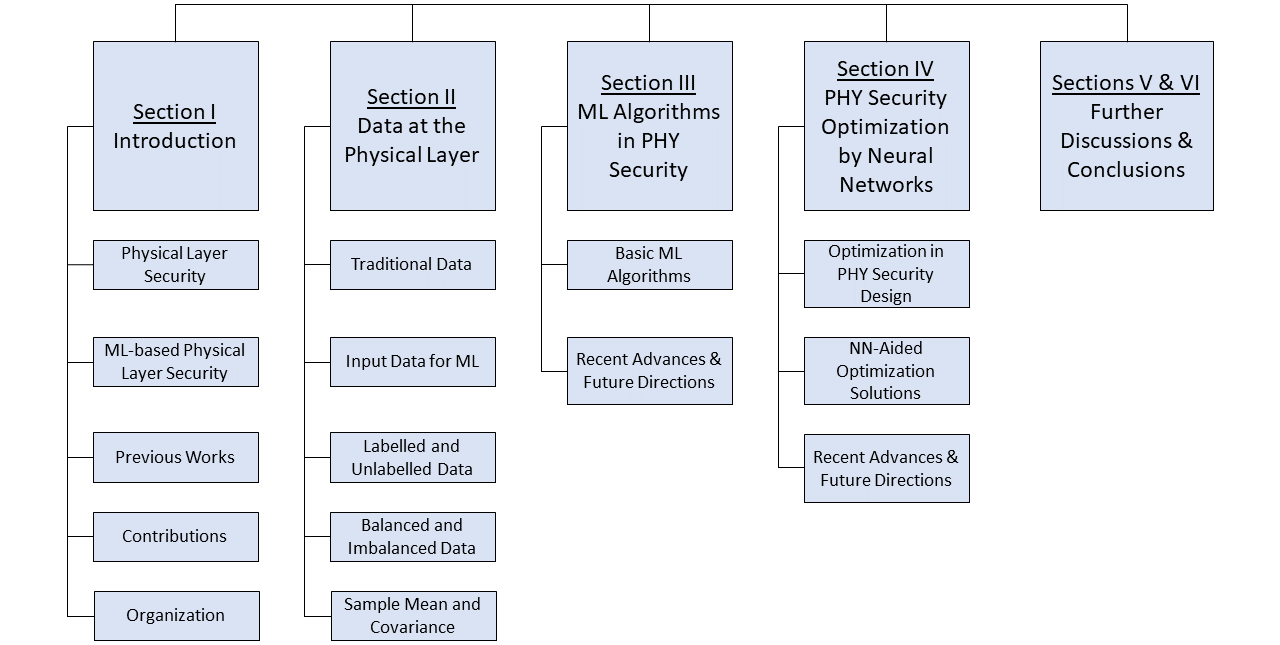}}
	\caption{The structure of the paper.}
	\label{fig: overview}
\end{figure*}

The remainder of this survey is organized as follows. In Section \ref{SEC: DATA}, we present the typical types of data at the physical layer {training data} that can be beneficially used by ML algorithms. Section \ref{SEC: Supervised - CLASSIFICATION} presents typical supervised ML classification algorithms, and underlines their applications in PHY security. Similarly, Section \ref{SEC: Unsupervised - CLASSIFICATION} presents typical unsupervised ML classification algorithms, and discusses their applications in PHY security. The basis of neural networks is presented in Section \ref{SEC: NNs - CLASSIFICATION}. The application of neural networks for optimizing security designs is presented in Section \ref{SEC: Design - NN}. Future {research ideas and the lessons learnt} are discussed in Section \ref{sec: future} and Section \ref{sec: conclusion}, respectively. Fig.~\ref{fig: overview} shows the structure of this paper.

\begin{table*}[!t]
	\centering
	\caption{Table of Contents}
	\renewcommand{\arraystretch}{1.5}
	\begin{tabular}{|c|l|c|}		
\hline 
\textbf{SECTIONS}
& \multicolumn{1}{c|}{\textbf{MAIN CONTENTS}}
& \multicolumn{1}{c|}{\textbf{NOTES}}
\\ \hline \hline 
I
& \textbf{Introduction} 
& {  }
\\ \hline 
II 
& \textbf{The Data at the Physical Layer} 
& {  }
\\ \hline 
\multirow{6}{*}{ III }
& \textbf{ML Algorithms in PHY Security}
& {  } 
\\ \cline{2-3}
{  }
& \hspace{0.5cm} - Supervised Learning
& \multirow{3}{*}{ The basic principles of typical ML algorithms}        
\\ \cline{2-2}
{  }
& \hspace{0.5cm} - Unsupervised Learning  
& {  }    
\\  \cline{2-2}
{  }
& \hspace{0.5cm} - (Artificial) Neural Networks  
& {  }  
\\ \cline{2-3}
{  }
& \multirow{2}{*}{\hspace{0.5cm} - Recent Advances \& Future Directions}
& \multirow{2}{*}{ ML-aided PHY Authentication \& Security Design }
\\ 
{  }
& {  }
& { } 
\\ \hline  
\multirow{5}{*}{IV}
& \textbf{PHY Security Optimization with Neural Networks} 
& {   }
\\ \cline{2-3}
{  }
& \hspace{0.5cm} - Optimization Problems Arisen in PHY Security Design  
& \multirow{2}{*}{ Optimization Formulation }
\\ \cline{2-2}
{  }
& \hspace{0.5cm} - Optimization Problems Solved by Neural Networks 
& {  }
\\ \cline{2-3}
{  }
& \multirow{2}{*}{\hspace{0.5cm} - Recent Advances \& Future Directions} 
& \multirow{2}{*}{ Optimizing PHY Security Designs }
\\ 
{ }
& { }
& { }
\\ \hline 
V
& \textbf{ML-aided PHY Security for Future Communication Systems}
& Further Discussion
\\ \hline
\end{tabular}
\label{table of contents}
\end{table*}
	
%
%
%
%
%
%
%
%
%

\section{Data at the Physical Layer}\label{SEC: DATA}
{Let us now briefly present} the traditional types of data that are routinely considered at the physical layer {followed by highlighting conversion of} the traditional data into the input data to be fed into ML algorithms {for their training}. 
\subsection{Traditional Data}\label{subsec: wireless data}
\begin{figure}[!t]	
\centerline{\includegraphics[width=1\linewidth]{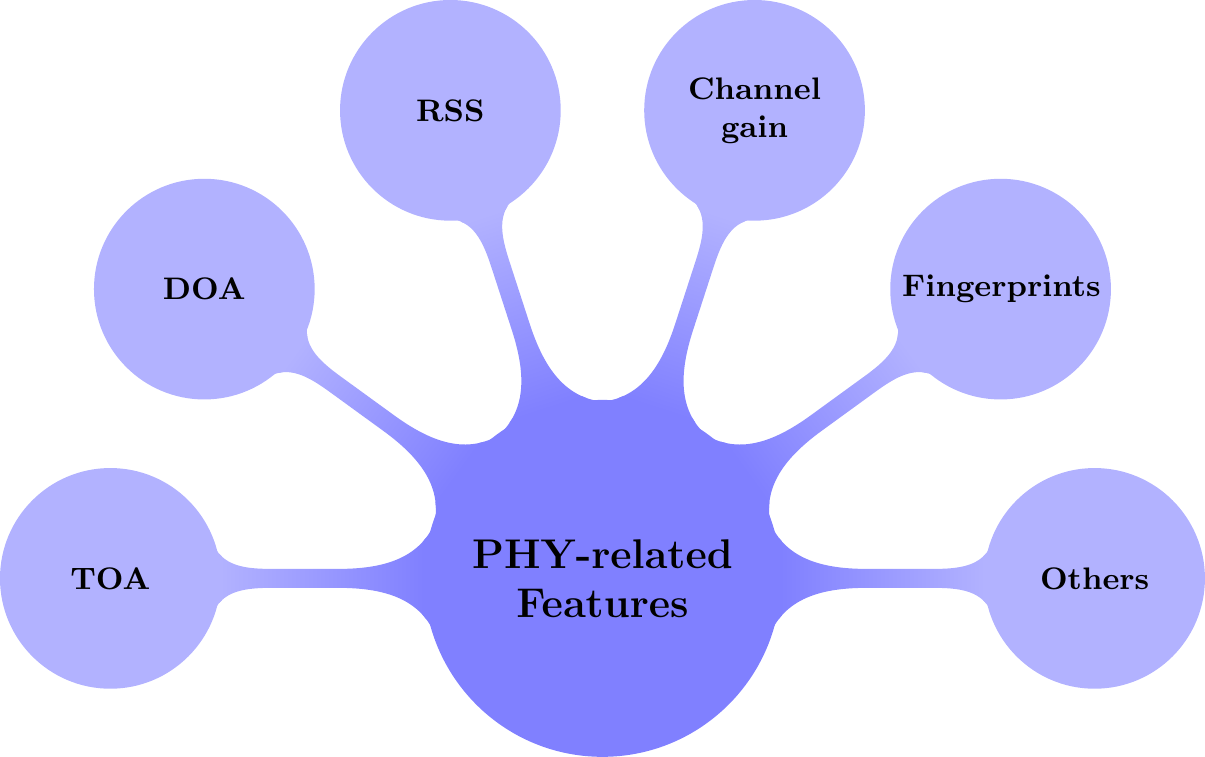}}
	\caption{Potential features for PHY security issues.}
	\label{fig: features}
\end{figure}

This subsection presents several typical features of the data {handled by} communication systems. Figure \ref{fig: features} provides an overview of the potential features {evaluated} for PHY authentication, including five specific types of data that are presented in this subsection. 
\subsubsection{Time of Arrival (TOA)}
The TOA is a common metric used for characterizing the time dispersion of a signal. We can express the ToA (in seconds) as follows:
\begin{align}
	t = t_0 + d/c + n + e_{\textrm{NLoS}} ,
\end{align}
where $t_0$ (s) is the transmit time {instant}, $d$ (m) is the distance between the transmitter and receiver, $c=3\times 10^8 (\textrm{m}/\textrm{s})$ is the speed of light, $n$ is the noise, and $e_{\textrm{NLoS}} \geq 0$ is the measurement error mainly caused by non-line-of-sight (NLoS) propagation. In general, it is widely accepted that $n$ is Gaussian distributed. For the NLoS error, several distributions have been used {for characterizing} the impact of NLoS conditions on time dispersion. For example, $e_{\textrm{NLoS}}$ can be described by an exponentially distributed random variable \cite{NLoS_ToA-AoA-RSS, ToA_Vaghefi_ExpDistributed, ToA_Guvenc_ExpDistributed}, a Gaussian distributed random variable \cite{ToA_GaussianDistributed}, or a uniformly distributed random variable \cite{ToA_UniformlyDistributed}. Note that $e_{\textrm{NLoS}}=0$, if there is no NLoS propagation. 

%
%

\subsubsection{Direction of Arrival (DOA)}
Another type of data that a receiver can attain is the DOA of a certain signal. The DOA estimation can be performed using direction-finding methods, such as the popular multi-signal classification (MUSIC) algorithm and Capon's minimum variance method \cite{DOA_MUSIC_Algorithm_BOOK, DOA_Dey2018}. 
In general, the performance of the DOA estimation will much depend on the receiver antenna configuration. Let $\B{a}$ be the array response in a direction $\theta$ and $\{\B{y}_1, \ldots,\B{y}_K \}$ be a set of received signals. The estimated value of $\theta$ can be found using the MUSIC algorithm as follows \cite{AoAEstimation-MUSIC-1, TMHoang2022-AOA-MUSIC-ML-AANET}:
\begin{align}
	\theta = \arg \max \frac{1}{ \left| \B{a}(\theta)^H \B{U} \right|^2}
\end{align}
where $(\cdot)^H$ denotes the conjugate transpose operator, and $\B{U} \B{\Sigma} \B{U}^H = \frac{1}{K} \sum_{k=1}^{K} \B{y}_k \B{y}_k^H$.
 
The DOA estimation plays a {pivotal} role in many applications such as positioning and tracking systems. From the PLS perspective, the DOA estimation techniques are also expected to play an important role in localizing the eavesdroppers' positions, or at least the eavesdropper's direction. 

\subsubsection{Received Signal Strength (RSS)}
The RSS can be {readily} quantified through the measurement of the received power at a certain receiver. More specifically, if $P_r(d)$ is the \emph{empirically} received power at a point $d$ meters away from the transmitter, then $P_r(d)$ can be viewed as a quantitative value for RSS at distance $d$. 
In practice, RSS values are likely to be inferred indirectly from the \emph{received signal strength indicator} (RSSI) values. It should be noted that RSSI values {represent} the readings obtained directly from a receiver. Moreover, the dynamic range of RSSI depends on {the quality of the} electronics manufacturers. Thus, if a simple receiver records the RSSI readings, {this has to be} translated from these values to the RSS values \cite{0.5k_RSS_RSSI_Clear_Relationship, RSSI_maps_to_actual_RSS}. For example, if TelosB motes\footnote{Each TelosB mote is integrated with an IEEE 802.15.4-compliant RF transceiver.} are used for measuring the RSSI values, then the corresponding RSS values can be calculated by using the following relationship \cite{0.5k_RSS_RSSI_Clear_Relationship}:
\begin{align}
	P_r(d) = \textrm{RSSI}(d) + \textrm{RSSI}_{\textrm{offset}}
\end{align} 
where $P_r(d)$ is in dBm, $\textrm{RSSI}(d)$ is the measured RSSI value (in dBm) at the distance $d$, and $\textrm{RSSI}_{\textrm{offset}} = -75$ dB is an offset value.

\subsubsection{Channel Gain}
Communication channels are random in nature and vary with the surrounding environment. Let us denote the channel between a transmitter and a receiver by $\B{h}_{channel}$. The role of the channel in a simple received signal model can be {formulated as} 
\begin{align}
	r_{Rx} = \B{h}_{channel}^{\dagger} \B{s}_{Tx} + n_{Rx} ,
\end{align}
where $\B{s}_{Tx}$ is the signal transmitted by multiple antennas, $r_{Rx}$ is the  signal received by a single antenna, $n_{Rx}$ is the receiver noise, and $^{\dagger}$ denotes the Hermitian operator.
Normally, $\B{h}_{channel}$ is assumed to obey a certain complex Gaussian distribution, but it depends on the type of channel fading. Common types of the channel fading are {modelled by} Rayleigh, Rician, and Nakagami-$m$ fading, just to name a few \cite{geolocation_NLoS_1, geolocation_2}. The {unique} random nature of $\B{h}_{channel}$ {is actually} helpful in dealing with eavesdropping, because secure transmission can be performed at the physical layer \cite{Survey_Security_Multi-Antenna, Zou2016}. 

\subsubsection{Fingerprints}
Hardware imperfections during the manufacturing process are unavoidable. However, that makes each hardware device unique and thus, the uniqueness of hardware is what can also be exploited for device identification \cite{Survey_Fingerprints_Baldini2017, PHY_authentication_Manufactoring_Process_2019}. The small differences in hardware are often termed as hardware \emph{fingerprints}. From a security point of view, hardware fingerprints {constitute} unique characteristics that can be exploited as the {unique} features of the input data for ML algorithms. The typical fingerprints include local oscillator frequency  \cite{PHY_authentication_Manufactoring_Process_2019}, carrier frequency offset \cite{Carrier_Frequency_Offset_1}, and the mismatch between in-phase and quadrature-phase components \cite{in-phase}. 

\subsection{Input Data for ML}
Once the wireless data {to be evaluated} has been collected, it is necessary to process the data to make them more useful. If wireless data is viewed as the raw data {before preprocessing}, then the processed data may be fed into ML models. The process of transforming the raw data into the processed data is termed as data preparation {required for} efficient ML-based {processing}.   
\begin{example}
Assume that we choose a certain channel vector $\textbf{h}_{\textrm{channel}} = \left[ 2+1j, 3-2j\right]^T$ to be the raw data and want to enter it into an ML model, but this model only accepts real-valued vectors as its input. Then, we  have to {find} a process that can create a new representation of {the} data, namely $\textbf{x}_{\textrm{input}}$, from the raw data $\textbf{h}_{\textrm{channel}}$. 
If we {opt for a} process that converts each element of $\textbf{h}_{\textrm{channel}}$ into its absolute square, then we will obtain $\textbf{x}_{\textrm{input}} = \left[ |2+1j|^2, |3-2j|^2\right]^T$. As a result, the processed data now meets the requirement of the ML model for real-valued input. 
\end{example}
ML algorithms are expected to work well on the processed data and produce good performance. In general, data preparation is the first stage before being able to apply ML algorithms to practical tasks. In data preparation, the following steps should be taken into account:
\begin{itemize}
    \item \textbf{Data cleaning}: Datasets from practical problems are not always {readily} available and properly formatted. Hence, there is a need to consider data cleaning so as to cope with missing data and transform categorized data into a numerical representation, which may be understandable {by} an ML model \cite{Data-clearning-Chu2016, Data-cleaning-2006}.
    \item \textbf{Feature selection}: When a  dataset contains irrelevant and redundant features, a process of selecting suitable features should be {harnessed as part of} a feature selection method in order to eliminate the irrelevant and redundant features \cite{Khalid2014, Zhao-Feature-Section}. These feature selection methods are also {often} referred to as search methods. {A range of} popular search algorithms can be used for performing feature selection, such as the exhaustive search and the branch-and-bound algorithm \cite{Meyer-Baese2014}.  
    \item \textbf{Feature extraction}: Another step to take is the process of extracting (or creating) new features from the original features. Depending on what type of data and problem we are handling, a suitable method of feature extraction will have to be used. In general, the new features conceived by feature extraction will have lower complexity than the original ones. For example, principal component analysis \cite{PCA-5640135} can transform the data from a high-dimensional space into a lower-dimensional space, while retaining as much valuable information as possible \cite{Khalid2014, Yuan-Feature-Extraction}.  
\end{itemize}

When it comes to the features of a dataset in a secure wireless system, it is still an open question, because there are no standards for defining the features. In general, the features can be extracted from information, {such as the} location, channel gain, or hardware \cite{FeatureExtraction-PLA5G-WangNing}. It is also possible to employ the metrics mentioned in Subsection \ref{subsec: wireless data} in order to create relevant features.

To illustrate what has been discussed above, let us consider a wireless system whose transmission relies on time slots. A receiver of the system considered might {rely on the} DOA and RSS to create a dataset. Let us denote the realization of DOA and that of RSS at the $t$-th time slot by $\textrm{DOA}[t]$ and $\textrm{RSS}[t]$. By arranging $\textrm{DOA}[t]$ and $ \textrm{RSS}[t]$ in a column of two elements, the receiver can form a data point as follows:
\begin{align}
\textbf{x}^{[t]}_{\textrm{input}} = 
\Bigl[ \Bigr.
\underbrace{ \textrm{DOA}[t] }_{\textrm{feature~} 1}, 
~
\underbrace{
\textrm{RSS}[t] 
}_{\textrm{feature~} 2}
\Bigl.\Bigr]^T
\nonumber 
\end{align}
in the two-dimensional space. 
In practice, $\textrm{DOA}[t]$ and $\textrm{RSS}[t]$ can take negative values, e.g., $\textrm{DOA}[t] = -45^\circ$ and $\textrm{RSS}^{[t]} = -20$ dB. If an ML model requires every element of $\textbf{x}^{[t]}_{\textrm{input}}$ to be positive, then we have to define $\textbf{x}^{[t]}_{\textrm{input}}$ in another way. For example, we can define it either as:
\begin{align}
\textbf{x}^{[t]}_{\textrm{input}} = 
\Bigl[ \Bigr.
\underbrace{ |\textrm{DOA}[t]|
}_{\textrm{feature~} 1},
~ 
\underbrace{
|\textrm{RSS}[t]| 
}_{\textrm{feature~} 2}
\Bigl. \Bigr]^T
\nonumber 
\end{align}
or as:
\begin{align}
\textbf{x}^{[t]}_{\textrm{input}} = 
\Bigl[ \Bigr.
\underbrace{ |\textrm{DOA}[t]|
}_{\textrm{feature~} 1},
~ 
\underbrace{
10^{ \textrm{RSS}[t] / 10 } 
}_{\textrm{feature~} 2}
\Bigl. \Bigr]^T.
\nonumber 
\end{align}
In doing so, $\textbf{x}^{[t]}_{\textrm{input}}$ can now be fed into that ML model. 
For the sake of generalization, let us define $M$ as the number of features. {Furthermore, let us} denote by $\Phi_m$ (with $m\in\{1, 2, \ldots, M\}$) the $m$-th feature of the data, where $\phi_m^{[t]}$ represents the observed value of $\Phi_m$ at the $t$-th time slot.
After $T$ time slots, the receiver will have $T$ data points that form the data input {formulated as}
\begin{align}
\textbf{X}_{\textrm{input}} 
&=
\left[
  \begin{array}{cccc}
    \vertbar & \vertbar & & \vertbar
    \\ \vspace{0.2cm}
    \textbf{x}^{[1]}_{\textrm{input}} & \textbf{x}^{[2]}_{\textrm{input}} & \ldots & \textbf{x}^{[T]}_{\textrm{input}} 
    \\
    \vrule & \vrule & & \vrule 
  \end{array}
\right]
\nonumber \\ 
&=
\begin{blockarray}{*{4}{c} l}
\begin{block}{[*{4}{c}]>{$\footnotesize}l<{$}}
  \phi_1^{[1]} &\phi_1^{[2]} &\ldots &\phi_1^{[T]} \bigstrut[t]& feature $\Phi_1$
  \\
  \phi_2^{[1]} &\phi_2^{[2]} &\ldots &\phi_2^{[T]} & feature $\Phi_2$ 
  \\
  \vdots &\vdots	&\ddots &\vdots & ~~~\vdots
  \\
  \phi_M^{[1]} &\phi_M^{[2]} &\ldots &\phi_M^{[T]} & feature $\Phi_M$ 
  \\
\end{block}
   \begin{block}{*{4}{>{$\footnotesize}c<{$}} l}
      slot 1 & slot 2 & \ldots & slot T & \\
   \end{block}
\end{blockarray} .
\label{X_input}
\end{align}

To generalize what will be discussed in the rest of the paper, we will not restrict the number of features. This means that the number of elements in the column vector $\textbf{x}_{\textrm{input}}^{[t]}$ can be an arbitrary positive integer. Moreover, it is not necessary to employ the DOA and RSS to create features. Instead, the feature selection and extraction will depend on the particular problems considered.

\subsection{Labelled and Unlabelled Data}
In authentication problems, it is important to recognize the presence and intrusion of eavesdroppers. Once an eavesdropper has been detected, we can stick a label {on the specific} data input that is related to that eavesdropper. If we stick the label $y_{\textrm{label}}^{[t]} = (-1)$ on the input data $\textbf{x}_{\textrm{input}}^{[t]}$ in order to mark a wireless system {pertubed by} eavesdroppers, then we may stick another label, e.g., $y_{\textrm{label}}^{[t]} = (+1)$ {on a secure} wireless system {free} any eavesdropper. Based on the availability of $\textbf{x}_{\textrm{input}}^{[t]}$ and $y_{\textrm{label}}^{[t]}$, three types of data can be formed:
\begin{itemize}
    \item \textbf{Labelled data}: From the statistical knowledge of previous transmissions, a wireless system can learn about eavesdroppers. For example, if the system suspects the presence of eavesdroppers at a certain time slot $t$ (with $t\in\{1,\ldots, T\}$), then it will be able to form a labelled dataset:
    \begin{figure}[!h]
    \centering
    \begin{tabular}{|c||c|c||}
    \hline \cline{2-3}
    Time  &\multicolumn{2}{c||}{Labelled data} \\
    \cline{2-3}
    slot & Input    & Output   \\
    \hline
    1 &$\textbf{x}_{\textrm{input}}^{[1]}$      
        &$y_{\textrm{label}}^{[1]}$         \\
    \vdots & \vdots & \vdots      \\
    $T$ &$\textbf{x}_{\textrm{input}}^{[T]}$       
        &$y_{\textrm{label}}^{[T]}$      \\
    \hline \cline{2-3} 
    \end{tabular}
    \end{figure}
    \vspace{0.1cm}
    There are two possibilities: 
    \begin{enumerate}
        \item If $y_{\textrm{label}}^{[t]}$ is the same for $\forall t\in\{1,\ldots, T\}$, then the labelled dataset has only a single class (or one label). 
        \item {However, if we have} $y_{\textrm{label}}^{[t]} \neq y_{\textrm{label}}^{[t']}$, with $t\neq t'$ and $t',t\in\{1,\ldots, T\}$, then the labelled dataset has two classes (or two labels).
    \end{enumerate}
    \item \textbf{Unlabelled data}: When there is no information about any eavesdropper and we cannot {ascertain whether} the wireless system is secure, then the value of $y_{\textrm{label}}^{[t]}$ at the $t$-th time slot is simply unknown to the system. {Explicitly, we must not put} the \emph{secure label} $(+1)$ on the output $y_{\textrm{label}}^{[t]}$, if we cannot be sure whether or not the corresponding input $\textbf{x}_{\textrm{input}}^{[t]}$ is related to eavesdroppers. As a result, unlabelled data will not {produce any} data output.
\end{itemize}

In practice, it is difficult to attain any knowledge {about the covert} eavesdroppers. For example, a legitimate link between a legitimate user and a receiver may be perfectly known to the receiver. However, {the illegitimate} link between an eavesdropper and the receiver considered may be {completely} unknown {since a so-called passive eavesdropper never transmits and hence, its channel cannot be estimated}. Thus, it is reasonable to consider a practical wireless network {to have no} channel state information (CSI) {for any of} the eavesdroppers. This means that real-world datasets are likely to contain only {a single} class, i.e., $(+1)$. {The lack of any knowledge about} the eavesdroppers' CSI {hampers the associated security analysis and design}. Hence, typically, the assumption of having \emph{imperfect} CSI for the eavesdroppers is used \cite{Hoang2020_WCL}. {Moreover, further research} is conducted to {develop sophisticated methods of detecting passive eavesdroppers} \cite{Detect-passive-Eve}.

\subsection{Balanced and Imbalanced Data}\label{subsec: data: imbalanced}
Let us {now return} to the labelled data and partition it into two groups: one of them includes all outputs $(+1)$, while the other includes all outputs $(-1)$. {Let us now} denote by $T_{(+1)}$ and $T_{(-1)}$ the number of data points in the first group and in the second group, respectively. {Naturally,} $T_{(+1)} + T_{(-1)} = T$ is the total number of data points that have been collected and labelled so far. Depending on the ratio $r_{\textrm{freq}} = \frac{T_{(-1)}}{T}$, the data can be considered as balanced or imbalanced. If this ratio is nearly $0.5$, the data is considered as balanced, since the number of data points in each class is nearly the same. By contrast, if the ratio $r_{\textrm{freq}}$ {tends to} $0$ (or $1$), the data becomes imbalanced. The value of $r_{\textrm{freq}}$ {informs us of} how frequently eavesdroppers {attempt to perturb} the system \cite{Hoang2020_WCL}.

Let us define 
\begin{align}
\mathcal{T}_1 &= \left\{ t| y_{\textrm{label}}^{[t]} = (+1) \right\} ,
\label{T_1}
\\
\mathcal{T}_2 &= \left\{ t| y_{\textrm{label}}^{[t]} = (-1) \right\} = \{1,\ldots,T\} \setminus \mathcal{T}_1 .
\label{T_2}
\end{align}
As such, the number of elements in $\mathcal{T}_1$ is $| \mathcal{T}_1 | = T_{(+1)}$, while that number in $\mathcal{T}_1$ is $| \mathcal{T}_2 | = T_{(-1)}$. Fig. \ref{fig: imbalanced data table} presents an example in which eavesdroppers {are present in} a wireless system $| \mathcal{T}_2 |$ times during $T$ transmissions. Let us assume {for example} that $| \mathcal{T}_2 | = 10$ and $T=1000$. Then the ratio of $T_{(-1)}$ to $T$ is $r_{\textrm{freq}}=10/1000$. 
\begin{figure}[!h]
    \centering
    \begin{tabular}{|c||c|c||c|}
    \cline{1-3}
    Counting elements &\multicolumn{2}{c||}{Labelled data} &\multicolumn{1}{c}{} \\
    \cline{2-3}
    in each class & Input & Output  &\multicolumn{1}{c}{}  \\
    \hline
    1 &\vdots    &$(+1)$   &\multicolumn{1}{c|}{\multirow{4}{*}{
    \rotatebox[origin=c]{-90}{No attack}
    }}      \\
     \vdots &$\textbf{x}_{\textrm{input}}^{[\tau]}$ &\vdots & \\
     $|\mathcal{T}_1| = 990$ &\vdots  &$(+1)$ & \\
    \hline  
    1 &\vdots &$(-1)$ & \multicolumn{1}{c|}{\multirow{3}{*}{\rotatebox[origin=c]{-90}{Attack}}} \\
    \vdots & $\textbf{x}_{\textrm{input}}^{[\tau']}$ & \vdots  &   \\
    $|\mathcal{T}_2| = 10$ &\vdots  &$(-1)$ & \\
    \hline 
    \end{tabular}
    \caption{An example of imbalanced data table.}
    \label{fig: imbalanced data table}
\end{figure}
In Fig. \ref{fig: imbalanced data table}, the data point $\textbf{x}_{\textrm{input}}^{[\tau]}$, which corresponds to the $\tau$-th time slot, is classified as $(+1)$. {Furthermore,} $\textbf{x}_{\textrm{input}}^{[\tau']}$ is classified as $(-1)$, where $\tau'\neq\tau$.

As such, if eavesdroppers are present frequently in the case of passive eavesdropping (or they attack frequently in the case of active eavesdropper) and we are aware of them, then we may build balanced data in which {the cardinality of the} two groups of data points is nearly equal. However, if the eavesdroppers are not present frequently, we may have imbalanced data. 

For balanced data, we can use a wide range of ML algorithms to {distinguish} the two classes, which is a basic binary classification problem. Widely-used ML models include logistic regression, support vector machine (SVM), neural networks, random forests, and decision trees, just to name a few.

For imbalanced data, several types of ML algorithms can be used to distinguish the group with fewer data points from the group with more data points. Among those algorithms are the isolation forest \cite{Isolation_Forest} and one-class support vector machine (OC-SVM) \cite{Wang-2004-OneClassSVM}.

\subsection{Sample Mean and Covariance}\label{subsec: data: mean and covariance}
{Let us} denote the sample mean of the class $(+1)$ and that of the class $(-1)$ by $\overline{\B{x}}_{\mathcal{T}_1}$ and $\overline{\B{x}}_{\mathcal{T}_2}$, respectively. {Then,} we have
\begin{align}
\overline{\B{x}}_{\mathcal{T}_1} &= \frac{1}{|\mathcal{T}_1|} \sum_{t\in\mathcal{T}_1} \B{x}_{\textrm{input}}^{[t]} ,
\label{x_mean: T1}
\\
\overline{\B{x}}_{\mathcal{T}_2} &= \frac{1}{|\mathcal{T}_2|} \sum_{t\in\mathcal{T}_2} \B{x}_{\textrm{input}}^{[t]} .
\label{x_mean: T2}
\end{align}
The overall mean of the entire population (i.e., the set $\{ \B{x}_{\textrm{input}}^{[1}, \ldots, \B{x}_{\textrm{input}}^{[T]} \}$) is calculated as
\begin{align}
\overline{\B{x}} &= \frac{1}{T} \sum_{t=1}^T \B{x}_{\textrm{input}}^{[t]} 
= \frac{1}{T} \left( |\mathcal{T}_1| \overline{\B{x}}_{\mathcal{T}_1} + |\mathcal{T}_2| \overline{\B{x}}_{\mathcal{T}_2} \right) .
\label{x_mean for the whole training data}
\end{align}

Moreover, we can also calculate the sample covariance matrices for the class $(+1)$ and the class $(-1)$ as follows \cite[Ch.5]{Flury2013}:
\begin{align}
	\B{\Sigma}_{sample, 1} &= \frac{
		\sum_{t\in \mathcal{T}_1} 
		\left( \B{x}_{\textrm{input}}^{[t]} - \overline{\B{x}}_{\mathcal{T}_1} \right)
		\left( \B{x}_{\textrm{input}}^{[t]} - \overline{\B{x}}_{\mathcal{T}_1} \right)^T
	}
	{ \left( |\mathcal{T}_1| - 1 \right) }  ,
	\label{covariance: Sigma sample 1}
	\\
	\B{\Sigma}_{sample, 2} &= \frac{
		\sum_{t\in \mathcal{T}_2} 
		\left( \B{x}_{\textrm{input}}^{[t]} - \overline{\B{x}}_{\mathcal{T}_2} \right)
		\left( \B{x}_{\textrm{input}}^{[t]} - \overline{\B{x}}_{\mathcal{T}_2} \right)^T
	}
	{ \left( |\mathcal{T}_2| - 1 \right) } .
	\label{covariance: Sigma sample 2}
\end{align}

\section{ML Algorithms in PHY Security}
\begin{figure}[t!]	\centerline{\includegraphics[width=1\linewidth]{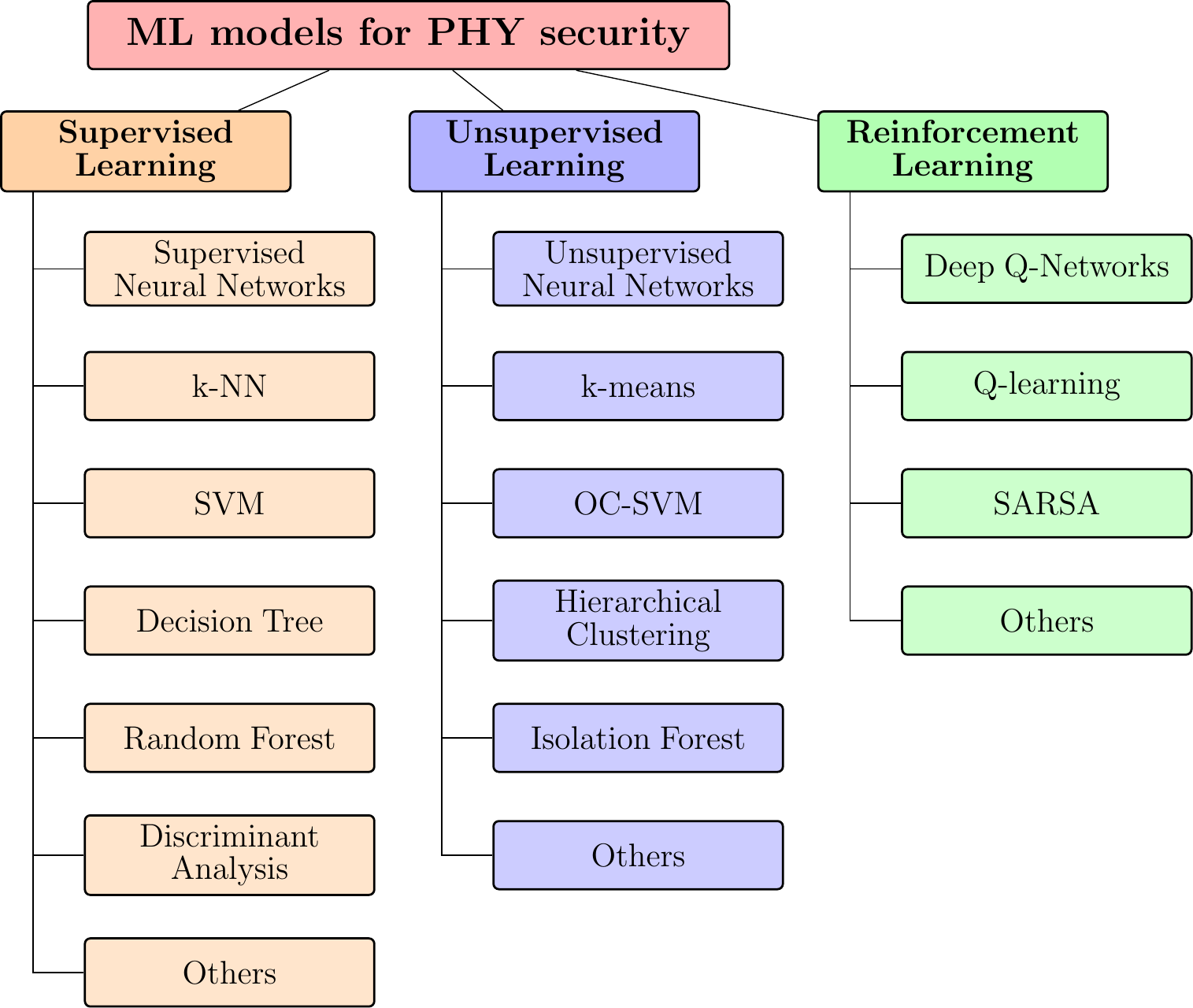}}
	\caption{The scope of this survey: Using ML models for detecting eavesdropping attacks and designing security strategies. A simplified anatomy of several ML models is also presented.}
	\label{fig: anatomy of ML}
\end{figure}

In this section, we review the basic principles of ML algorithms, including four typical supervised learning algorithms, four typical unsupervised learning algorithms, reinforcement learning, and neural networks. We then review as well as discuss the recent advances in applying ML algorithms to PHY authentication and security design. Figure \ref{fig: anatomy of ML} depicts the {family-tree} of ML algorithms.

\subsection{Basic ML Algorithms}
\subsubsection{Supervised Learning} \label{SEC: Supervised - CLASSIFICATION}
In this subsection, four typical supervised learning algorithms are summarized, {namely,} the $k$-nearest neighbour, support vector machine, random forest, and linear discriminant analysis.
\paragraph{K-Nearest Neighbour (k-NN)}
The k-nearest neighbour (k-NN) {clustering} belongs to the group of \emph{lazy learning} algorithms because it does not require any training process. Instead, it only uses its training data to classify new data points when those new data points are entered. Due to the lack of training, k-NN has to store all training data points for use, {which makes it} a computationally inefficient algorithm. 

\begin{figure}[!t]
	\centerline{\includegraphics[width=.6\linewidth]{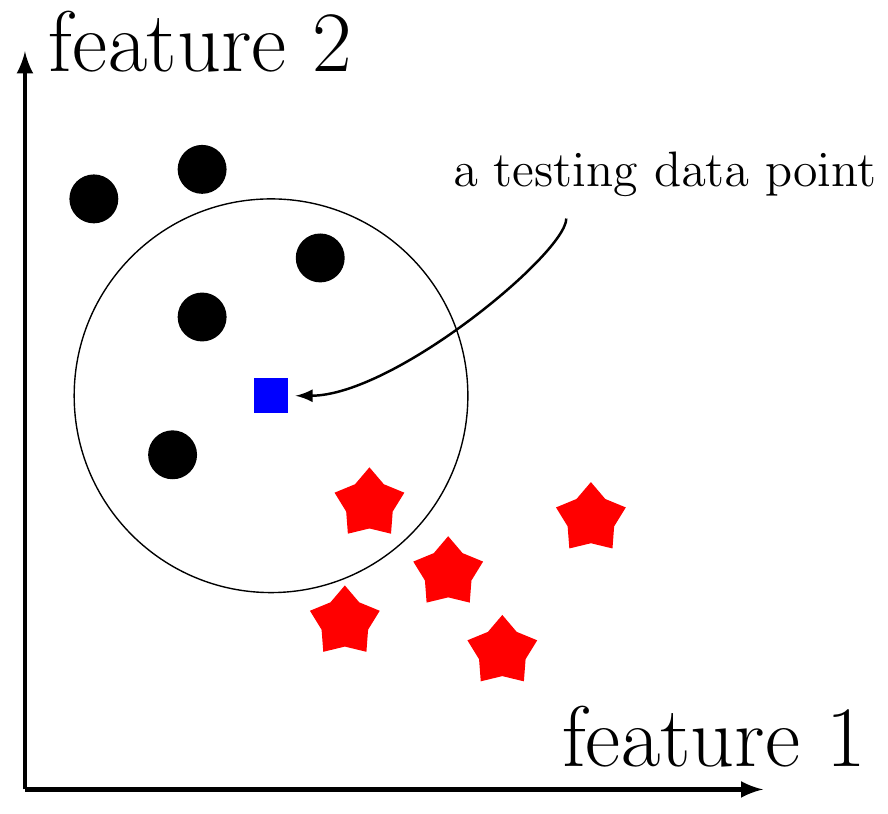}}
	\caption{An illustration of how $k$-NN classifies a new data point.}
	\label{fig: k_NN}
\end{figure}

In principle, when classifying a testing data point $\B{x}_{test}$, the $k$-NN algorithm considers the $k$ nearest neighbours of $\B{x}_{test}$ and then assigns $\B{x}_{test}$ to the class based on majority voting \cite{Bishop-2006-ML, Guo2003_k-NN, Garcia2008_k-NN, Liao2002_k-NN}. As illustrated in Figure \ref{fig: k_NN}, three out of four nearest neighbours of a testing instance belong to the class (+1), hence the testing instance will be classified as (+1). Since the $k$-NN algorithm processes all data points in the training data, it will become {particularly} inefficient when dealing with large datasets. 

\paragraph{Support Vector Machine (SVM)}
The SVM is a powerful type of ML, which can {efficiently} perform both linear and non-linear classification tasks when the size of the data {set} is small or medium. The goal of SVM is to find the optimal boundary that separates 2 different classes in a dataset. Soft-margin SVM offers more versatility than hard-margin SVM, because the soft-margin SVM accepts some misclassified data points, thus avoiding over-fitting. This means that the soft-margin SVM has the ability to generalize. Given that the equation of a hyperplane has the form $\hslash(\B{x}) = \B{x}^T \B{a} + b = 0$, the goal of SVM is find the optimal values of $\B{a}$ and $b$ through solving the following optimization problem:
\begin{subequations}\label{opt: P2}
	\begin{eqnarray}
	& \displaystyle \min_{ \B{a}, b,\xi_t } 
	&
	\underbrace{
		\frac{1}{2} \|\B{a}\|^2 
	}_{\text{regularizer}}
	+ 
	\underbrace{
		C 
		\sum_{t=1}^{T} \xi_t^{\kappa}
	}_{\text{error}}
	\label{opt: P2 a}\\
	&\text{s.t.}	
	& y_{\textrm{label}}^{[t]} \left( \B{a}^{\top} \B{x}_{\textrm{input}}^{[t]}   + b  \right) \geq 1 - \xi_t 
	\end{eqnarray}
\end{subequations}
where $C$ is the margin parameter {in \eqref{opt: P2 a}}. $\kappa=1$ implies the $1$-norm (L1) soft-margin SVM, while $\kappa=2$ implies the $2$-norm (L2) soft-margin SVM. 

\begin{figure}[!t]	\centerline{\includegraphics[width=1\linewidth]{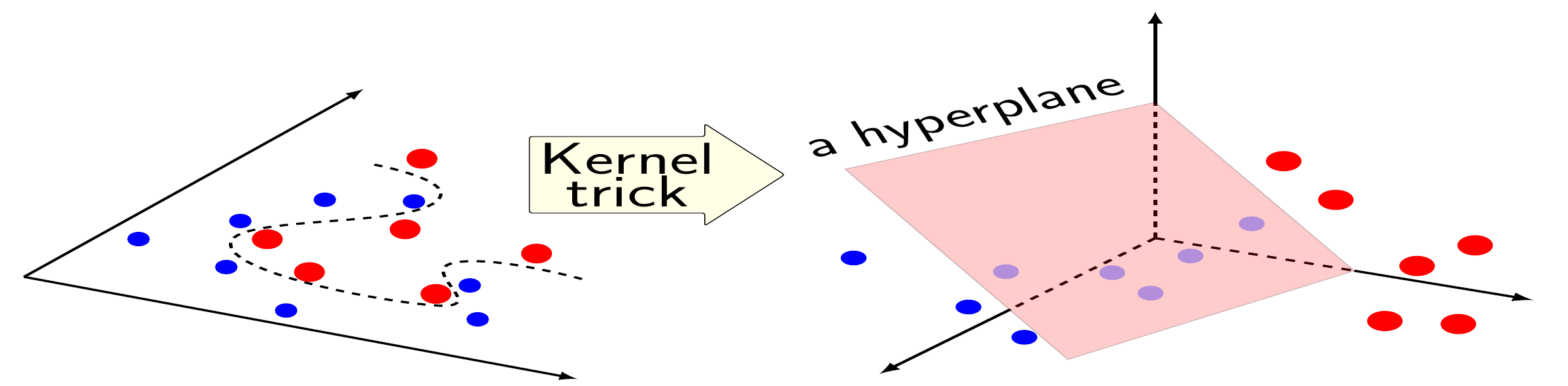}}
	\caption{The use of the kernel trick in SVM classifiers can bring the data to a higher-dimensional space, thereby making the data more separable.}
	\label{fig: SVM}
\end{figure}

When using SVM, a ``magic'' technique called \emph{kernel trick} {may be} exploited to transform the data from the original input space into a higher-dimensional space. The reason behind the use of the kernel trick is that the data in a higher-dimensional space may be linearly separable, {which was not separable} in the original space. Figure \ref{fig: SVM} illustrates how the kernel trick works, where a dataset with two features becomes more separable, when being transformed into a new space with three features.
   
Let $\phi(\cdot)$ be  a nonlinear mapping that stretches
the original input space $\mathcal{X}_{\textrm{input}}$ to some higher-dimensional space $\mathcal{H}$. For the $t$-th data point $\B{x}_{\textrm{input}}^{[t]}$, we have the corresponding value {of} $\phi_t = \phi\left( \B{x}_{\textrm{input}}^{[t]} \right)$. Then the inner product $\mathcal{K}(t,t') = \phi_t \phi_{t'}^{\top}$ is a kernel function. When using the kernel trick, we will encounter the calculation of $\mathcal{K}(t,t')$ instead of $\phi( \B{x}_{\textrm{input}}^{[t]} )$ and $\phi( \B{x}_{\textrm{input}}^{[t']} )$. This alleviates the computational cost because it is not necessary to calculate explicit expressions for $\phi( \B{x}_{\textrm{input}}^{[t]} )$ and $\phi( \B{x}_{\textrm{input}}^{[t']} )$. Moreover, kernel functions are normally predefined and there are many types of kernels to choose from, such as linear, radial basis function (RBF), polynomial, and sigmoid kernels.  

\begin{figure}[!t]
	\centerline{\includegraphics[width=.9\linewidth]{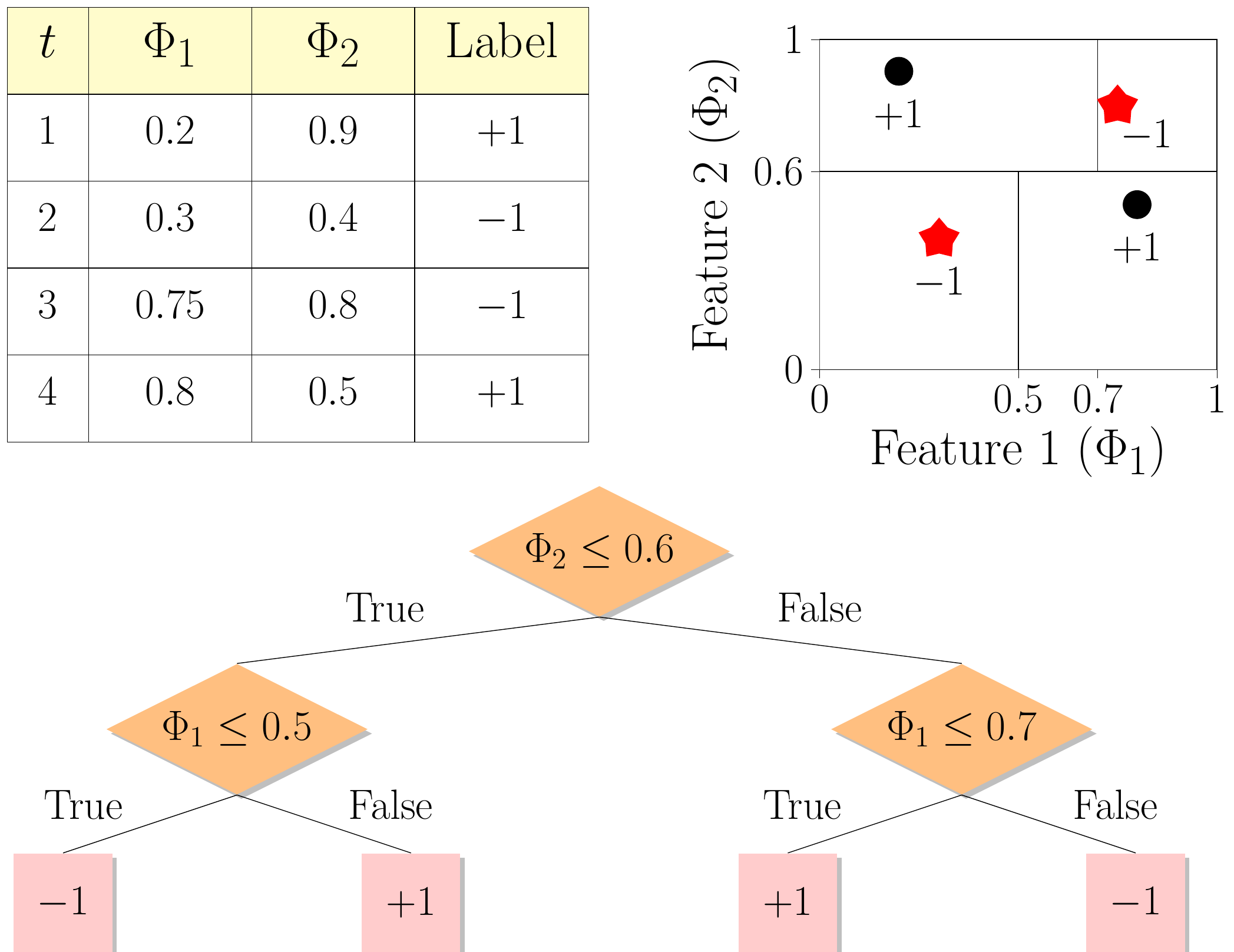}}
	\caption{An example of building a decision tree from a dataset.}
	\label{fig: DT}
\end{figure}

\paragraph{Random Forest}
A ``\emph{random forest}'' is basically a collection of decision trees. To understand the random forest algorithm, let us recall the basic idea behind the decision tree algorithm:

A decision tree is built top-down from a root node. The leaf nodes (i.e., the terminal nodes) of the tree represent decisions (i.e., labels). A decision node of the tree represents the test {of a specific} feature. Note that the topmost decision node is the root node. Each decision node has at least two branches each representing an outcome of the corresponding test. Figure \ref{fig: DT} depicts a decision tree {having} basic elements. 
To construct decision trees, we can use different algorithms. Among widely-used algorithms are ID3 (i.e., iterative dichotomiser 3), CART, CHAID, MARS and so on \cite{Assessment-of-Decision-Tree-2018, gupta2017analysis}. Let us consider the ID3 algorithm that uses entropy and information gain to construct a decision tree {based on measuring/scoring} the features of data. Based on the features that have been scored, a decision tree will be built top-down as follows: 
	\begin{itemize}
		\item Step 1: At a decision node, select the highest-score feature for a test.   
		\item Step 2: From that decision node, create branches and split the training data into subsets so that each subset corresponds to a branch.
		\item Step 3: Find leaf nodes and repeat {the steps} recursively on each subset.
	\end{itemize} 

To build a random forest from decision trees, we first need to create \emph{bootstrapped datasets} from the training data. By using a statistical technique called \emph{bootstrapping}, some data points in the training set will be randomly picked in order to create a bootstrapped dataset. Thus, a bootstrapped dataset is actually a subset of the training data. Note that the number of bootstrapped datasets is equal to the number of decision trees in a random forest. Moreover, a certain data point may belong to many bootstrapped datasets because the sampling allows a data point to be picked many times. This sampling method is also known as \emph{sampling with replacement}. Once the bootstrapped datasets have been prepared, we can build the decision trees each corresponding to a bootstrapped dataset. Then using \emph{majority voting} \cite{MajorVoting-RF-2014}, a new data point will be predicted.

\paragraph{Linear Discriminant Analysis (LDA)}
Linear discriminant analysis (LDA) falls {into the family of} discrimination techniques. {In contrast to} classification techniques that assign a new data point to a certain class, the purpose of discrimination is to make the data as separable as possible \cite[Ch. 11]{DiscriminantAnalysis_22k_Chapter11}. Linear discriminant analysis can also be used for binary classification and it will check if a received signal is associated with an attack. Let us assume that {a pair of} two samples belonging to labels $(+1)$ and $(-1)$ obeys two different statistical distributions with covariance matrices $\B{\Sigma}_{theory, 1}$ and $\B{\Sigma}_{theory, 2}$. In order to use linear discriminant analysis, it is required that $\B{\Sigma}_{theory, 1}$ is equal to $\B{\Sigma}_{theory, 2}$. In practice, $\B{\Sigma}_{theory, 1}$ and $\B{\Sigma}_{theory, 2}$ are normally unknown, thus it is reasonable to replace them with the so-called \emph{sample} covariance matrices $\B{\Sigma}_{sample, 1}$ and $\B{\Sigma}_{sample, 2}$, which are formed by the available samples in our training data (see Subsection \ref{subsec: data: imbalanced}) and are calculated in \eqref{covariance: Sigma sample 1}-\eqref{covariance: Sigma sample 2} in Subsection \ref{subsec: data: mean and covariance}. Accordingly, $\B{\Sigma}_{sample, 1}$ and $\B{\Sigma}_{sample, 1}$ are expected to be equal in order to use linear discriminant analysis. At this point, a problem arises from the fact that $\B{\Sigma}_{sample, 1}$ and $\B{\Sigma}_{sample, 2}$ may not be equal, i.e., $\B{\Sigma}_{sample, 1} \neq \B{\Sigma}_{sample, 2}$. To overcome this problem, we use a weighted average of $\B{\Sigma}_{sample, 1}$ and $\B{\Sigma}_{sample, 2}$, which is called the pooled sample covariance matrix. This matrix can be expressed as follows \cite[Ch.5]{Flury2013}, \cite{DiscriminantAnalysis_22k_Chapter11}:
\begin{align}
	\B{\Sigma}_{pooled} = 
	\frac{
		\left(|\mathcal{T}_1|-1\right) \B{\Sigma}_{sample, 1}
		+ \left(|\mathcal{T}_2|-1\right) \B{\Sigma}_{sample, 2} 
	}{ |\mathcal{T}_1| + |\mathcal{T}_2| - 2 },
\end{align}
where $|\mathcal{T}_1|$ and $|\mathcal{T}_2|$, defined in eqs.\eqref{T_1}--\eqref{T_2}, are the number of samples in the class $(+1)$ and that in the class $(-1)$, respectively. Then, we use $\B{\Sigma}_{pooled}$ as the single common covariance matrix for the two populations of interest.

\begin{figure}[!t]
	\centerline{\includegraphics[width=.95\linewidth]{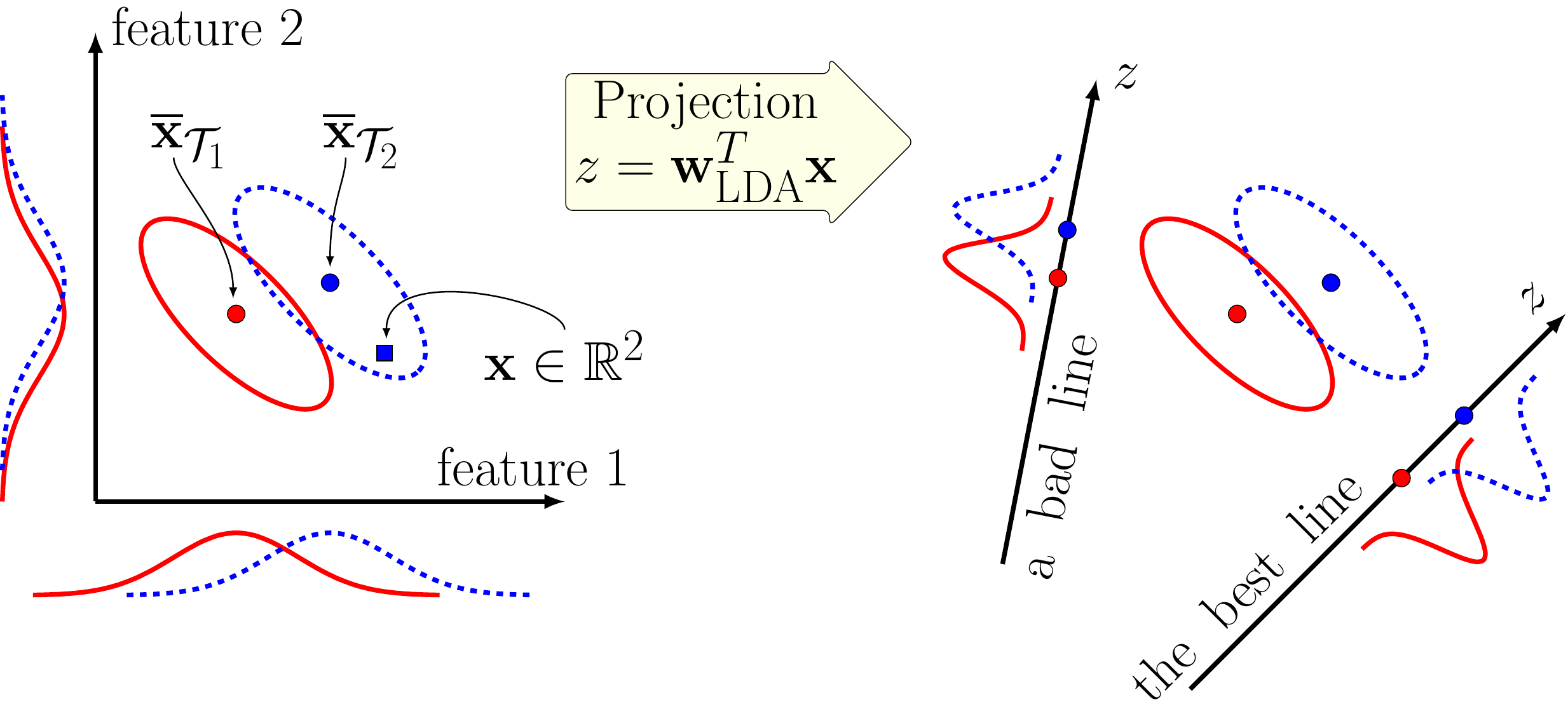}}
	\caption{An illustration of how linear discriminant analysis can help with separating a sample from the other. In this figure, we assume that the training data has two features. A new feature will be created to replace the old ones by using a projection onto a certain line. Using linear discriminant analysis allows us to find the best line $z=\B{w}_{\textrm{LDA}}^T \B{x}$. Each ellipse represents the confidence region of a distribution \cite{Wang2015, Friendly2013}.}
	\label{fig: LDA}
\end{figure}
In theory, linear discriminant analysis seeks some transformation vector $\B{w}_{\textrm{LDA}}$, which is used to project a data point $\B{x}$ in the input space onto a line $z = \B{w}_{\textrm{LDA}}^T \B{x}$, so that the function
\begin{align}
\lambda_{\textrm{LDA}} = 
\frac{  
\B{w}_{\textrm{LDA}}^T \B{S}_{\textrm{between}} \B{w}_{\textrm{LDA}}
}
{ 
\B{w}_{\textrm{LDA}}^T \B{S}_{\textrm{within}} \B{w}_{\textrm{LDA}} 
}
\label{LDA: objective func}
\end{align}
is maximized. In \eqref{LDA: objective func}, $\B{S}_{\textrm{within}}$ is the within-class (scatter) matrix, while $\B{S}_{\textrm{between}}$ is the between-class (scatter) matrix.\footnote{The between-class matrix is different from the between-class covariance matrix. To obtain the latter, we can divide the former by $|\mathcal{T}_1| + |\mathcal{T}_2| - 1$.} These matrices can be expressed as \cite{Konishi2014, Flury2013}
\begin{align}
\B{S}_{\textrm{within}} &= \left(|\mathcal{T}_1| + |\mathcal{T}_2| - 2\right) \B{\Sigma}_{pooled} ,
\\
\B{S}_{\textrm{between}} &= 
\sum_{j=1}^{2} 
|\mathcal{T}_j|
\left( \overline{\B{x}}_{\mathcal{T}_j} - \overline{\B{x}} \right)
\left( \overline{\B{x}}_{\mathcal{T}_j} - \overline{\B{x}} \right)^T,
\end{align}   
where $\overline{\B{x}}_{\mathcal{T}_1}$ and $\overline{\B{x}}_{\mathcal{T}_2}$ are the sample mean of the class $(+1)$ and that of the class $(-1)$, respectively. The expressions for $\overline{\B{x}}_{\mathcal{T}_1}$ and $\overline{\B{x}}_{\mathcal{T}_2}$ are are defined in \eqref{x_mean: T1} and \eqref{x_mean: T2}, while $\overline{\B{x}}$ is the sample mean of the whole training data and it is defined in \eqref{x_mean for the whole training data}.
Finding the maximal value of $\lambda_{\textrm{LDA}}$ leads to $\B{w}_{\textrm{LDA}} = \B{S}_{\textrm{within}}^{-1} \left( \overline{\B{x}}_{\mathcal{T}_1} - \overline{\B{x}}_{\mathcal{T}_2} \right)$. 
{Observe} from \eqref{LDA: objective func} that $\lambda_{\textrm{LDA}}$ is also an eigenvector of the matrix
$
\B{S}_{\textrm{LDA}} \mathop=\limits^{def}
\B{S}_{\textrm{within}}^{-1} \B{S}_{\textrm{between}}
.
$
Accordingly, linear discriminant analysis can be intuitively {interpreted} as a method in which the data points in the input data will be projected onto the \emph{largest} eigenvector of $\B{S}_{\textrm{LDA}}$. Figure \ref{fig: LDA} illustrates how linear discriminant analysis projects the two samples in the original $2$-dimensional space onto the largest eigenvector of $\B{S}_{\textrm{LDA}}$. 

By using linear discriminant analysis, a new data point $\B{x}_{new}$ in the test data will be classified as the first class if the inequality
\begin{align}
	&\left( \overline{\B{x}}_{\mathcal{T}_1} - \overline{\B{x}}_{\mathcal{T}_2} \right)^T \B{\Sigma}_{pooled}^{-1} \B{x}_{new} 
	\nonumber \\
	&\geq
	\frac{1}{2} \left( \overline{\B{x}}_{\mathcal{T}_1} - \overline{\B{x}}_{\mathcal{T}_2} \right)^T \B{\Sigma}_{pooled}^{-1} \left( \overline{\B{x}}_{\mathcal{T}_1} + \overline{\B{x}}_{\mathcal{T}_2} \right)
	+ \ln\left( \frac{ c(1|2) p_2 }{ c(2|1) p_1 } \right)
	\label{LDA: comparison}
\end{align}
holds. Herein, $c(1|2)$ and $c(2|1)$ {represent} the costs of misclassification, while $p_1$ and $p_2$ are the prior probabilities of the first class and the second class, respectively. It should also be noted that the left hand side of \eqref{LDA: comparison} is a point on the line, which the largest eigenvector passes through. {To ellaborate further}, the right hand side of \eqref{LDA: comparison} is a given threshold that is used for binary classification. Note that the principle of linear discriminant analysis can also be extended to multi-class classification \cite{Tharwat2017}, \cite[Ch.6]{Konishi2014}.


\subsubsection{Unsupervised Classification Algorithms} \label{SEC: Unsupervised - CLASSIFICATION}
In this subsection, four typical unsupervised learning algorithms are summarized, namely $K$-means clustering, one-class support vector machine, isolation forest, and hierarchical clustering.
\paragraph{$K$-Means Clustering ($K$-means)}
$K$-means clustering {acts} directly on the testing data without requiring a training process \cite[Ch. 9]{Bishop-2006-ML}. The {benefit} of $K$-means clustering is that it can cluster the data into $K$ clusters. Using $K$-means clustering for detecting eavesdropping attacks, we may only have to  classify the data into two clusters by setting $K=2$, {where} one cluster corresponds to eavesdropping attacks, and the other to no attack. 

In general, $K$-means clustering yields $K$ clusters each having a centroid. At the beginning, $K$-means may generate {its} centroids randomly and then updates the positions of the centroids iteratively. At each iteration, each data point will be assigned to the closest centroid. Once all data points have been assigned to {one of} the centroids, the newly-created clusters may or may not be different from the earlier clusters. If the new clusters are not the same as the old clusters, it is necessary to update the centroids. The update rule of a centroid is {based on calculating} the average value of all the data points in the corresponding cluster. After the centroids have been updated, the iterative algorithm moves on to the next iteration. The algorithm will continue {its iterations} until there is no change to the centroids or the maximum number of iterations is reached. 

The accuracy of $K$-means clustering relies heavily on the initialization of centroids. Thus, different initializations result in different accuracy levels. Instead of randomly generating the initial centroids, {advanced} techniques suggest selecting the positions of initial centroids more carefully. For example, $K$-means++ {exhibits} its superiority over the standard $K$-means clustering in terms of both speed
and accuracy \cite{k-means++}. In essence, the only difference between $K$-means++ and $K$-means clustering lies in the initialization, {but} the iterative procedure in $K$-means++ clustering is the same as the standard $K$-means clustering. To be more precise, $K$-means++ starts with choosing a centroid randomly before calculating the distances of points in the dataset from the selected centroid. Then the next centroids are found based on considering a functional relationship of those distances. A detailed description of $K$-means++ initialization is presented in \cite{k-means++}.

\begin{figure}[!t]
\centerline{\includegraphics[width=.9\linewidth]{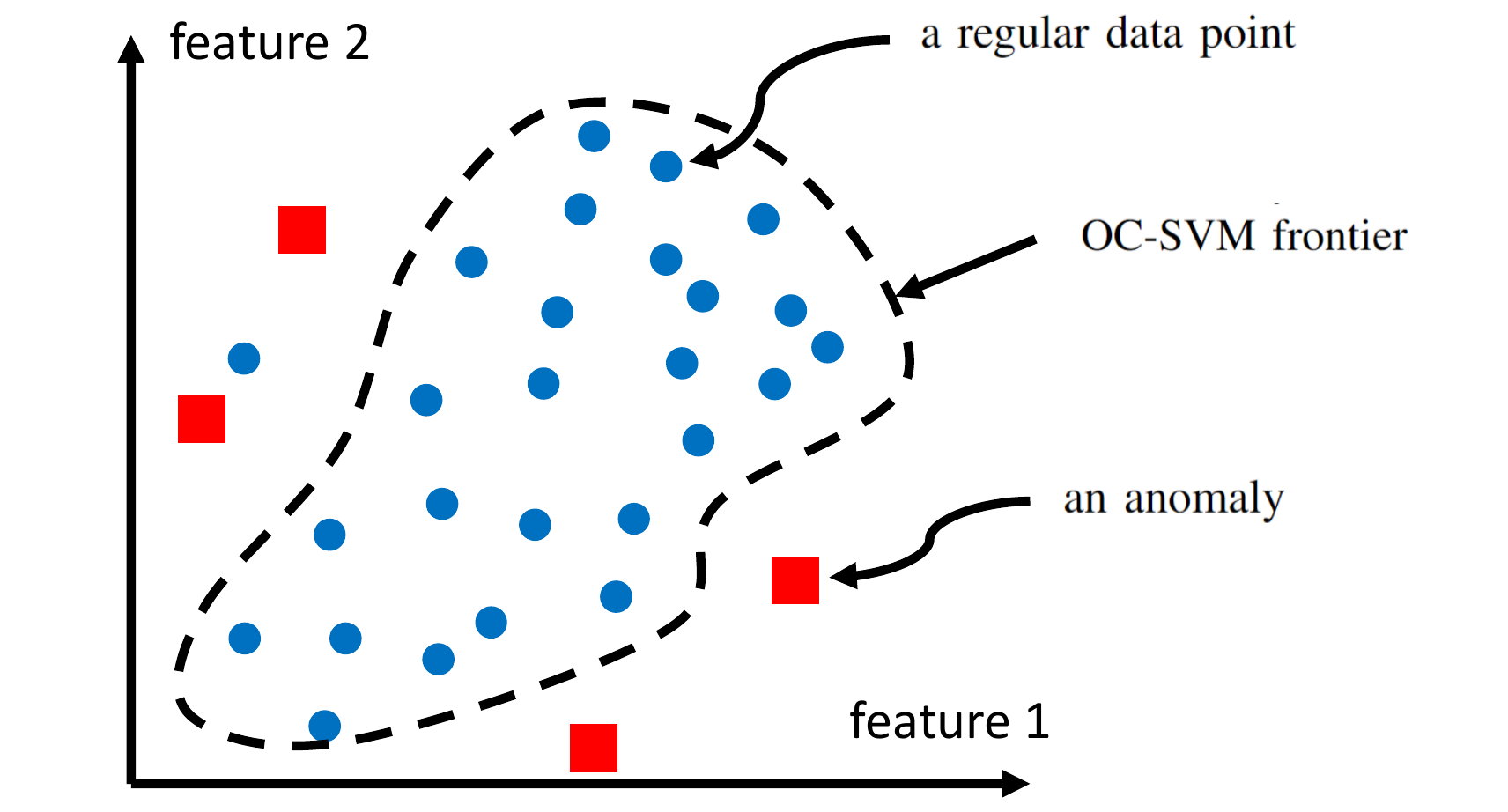}}	
	\caption{An illustration of the OC-SVM algorithm.}
	\label{fig: OC SVM 1}
\end{figure}

\paragraph{One-class Support Vector Machine (OC-SVM)}
As a variant of SVM, the one-class support vector machine (OC-SVM) has been developed to deal with outliers or imbalanced data. According to \cite{Manevitz-2001-OneClassSVM, Wang-2004-OneClassSVM, Scholkopf-2001-OneClassSVM}, the goal of OC-SVM is to {evaluate} the decision function 
\begin{align}
\hslash(\B{x}) = \mathrm{sign} \left( \B{w}^T \phi(\B{x}) - \rho \right) 
\end{align}
where $\mathrm{sign}(\cdot)$ is the sign function, $\phi(\cdot)$ is a mapping that stretches the input space $\mathcal{X}_{\textrm{input}}$ to a higher-dimensional space, {while} $\B{w}$ and $\rho$ are found by solving the optimization problem:
\begin{subequations}\label{Problem OC-SVM}
	\begin{alignat}{2}
	&
	\underset{
		\B{w}, \rho, \xi_i
	}{ \text{minimize} }
	\quad
	&& 
	\underbrace{
		\frac{1}{2} \|\B{w}\|^2
	}_{\textrm{regularizer}}
	+
	\underbrace{ 
		\frac{1}{\nu T_{tot}} \sum_{w=1}^{T_{tot}} \xi_w
		-\rho
	}_{\textrm{error}}
	\label{Problem P1: C1} \\
	&
	\text{subject to} \quad
	&
	&\langle \B{w}, \phi(\B{x}_{w}) \rangle \geq \rho - \xi_w .
	\end{alignat}
\end{subequations}
Herein, $\xi_w \geq 0$ is a slack variable, and $\nu\in(0,1]$ is a parameter balancing the maximal distance from the origin and the number of data points in the region created by the hyperplane \cite{Wang-2004-OneClassSVM}. 

In OC-SVM, any data points that cannot form a dense cluster will be treated as outliers/anomalies \cite{Scholkopf-2001-OneClassSVM}. In other words, the outliers reside within the regions of low density. In the meantime, regular data points will be treated as the inliers because they form the regions of high density. Figure \ref{fig: OC SVM 1} depicts the idea behind the use of OC-SVM.

To {highlight} how OC-SVM can be used in the context of PHY security, let us consider the case in which eavesdroppers rarely attack a wireless system and thus, the system can only use one-class (or one-label) data for training. In this case, OC-SVM can help with training the data and detect an outlier that is likely {to be} related to an eavesdropping attack. 

\begin{figure}[!tb]
	\centerline{\includegraphics[width=1\linewidth]{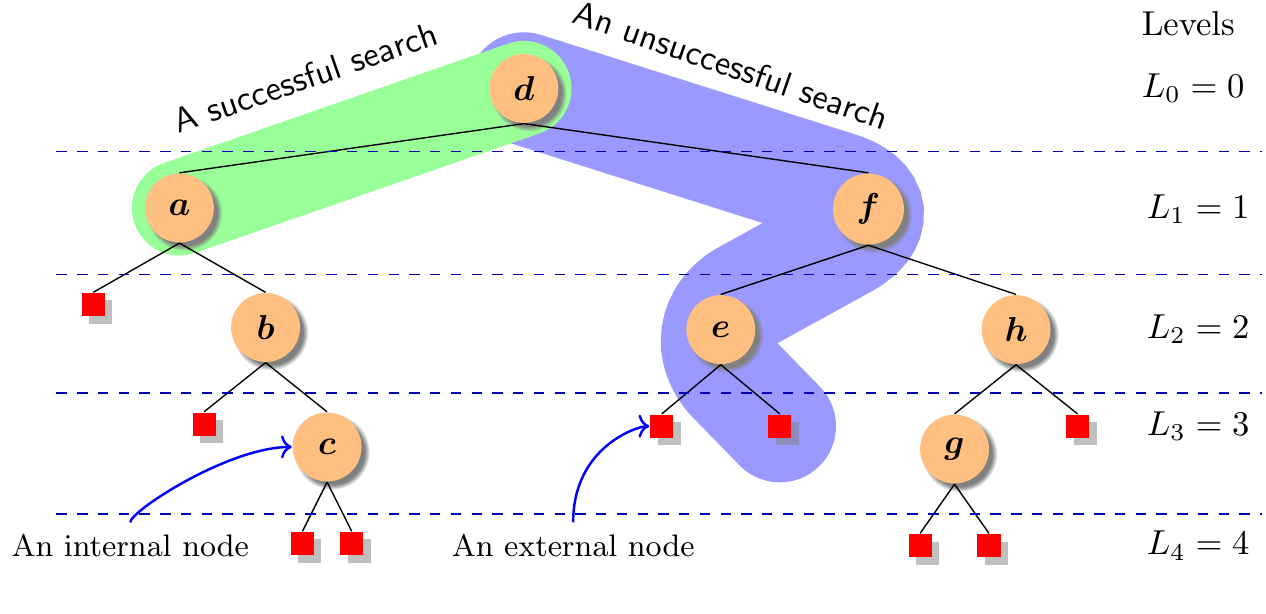}}
	\caption{An illustration of a binary search tree with $n_{int}=8$ internal nodes ($a<b<c<d<e<f<g<h$). While a successful search terminates at an internal node  (e.g., searching for $b$), an unsuccessful search terminates at an external node (e.g., searching for $x$ given that $e<x<f$). $\mathtt{IPL} =1 L_0 + 2L_1 + 3L_2 + 2L_3 + 0L_4 = 14$ and 
		$\mathtt{EPL} = \mathtt{IPL} + 2 n_{int} = 30$.
	}
	\label{fig: a BST}
\end{figure}
\paragraph{Isolation Forest}





An isolation forest is an ensemble method, whose anomaly scores are averaged over multiple isolation trees. In principle, the structure of an isolation tree is very similar to that of a binary search tree (BST). To understand the concept of an isolation tree, we first recall the binary search tree structure. 
\begin{remark}
	\emph{The binary search tree structure:} A binary search tree is a binary tree {having} the following properties: 
	\begin{itemize}
		\item Each node in a binary search tree has at most 2 child nodes.
		\item Each node in a binary search tree has a unique value (or key); denote the value of node $X$ by $\mathtt{Val}_X$. 
		\item Let $Y$ be a child node of $X$. If $\mathtt{Val}_Y < \mathtt{Val}_X$ (or $\mathtt{Val}_Y > \mathtt{Val}_X$), then $Y$ is the left (or right) sub-tree of $X$. 
	\end{itemize} 
\end{remark}
{Let us} denote the number of \emph{internal} nodes in a binary search tree by $n_{int}$, the \emph{internal} path length by $\mathtt{IPL}$, and the \emph{external} path length by $\mathtt{EPL}$. According to \cite{Binary_Search_Tree_Preiss1999}, we have $\mathtt{EPL} = \mathtt{IPL} + 2n_{int}$. By averaging over many binary search trees, the average external path length of a binary search tree is calculated as  $\overline{\mathtt{EPL}} = \overline{\mathtt{IPL}} + 2n_{int}$, where $\overline{\mathtt{IPL}}$ is the average internal path length of a binary search tree. {For} $n_{int}$ internal nodes, a binary search tree will have $n_{ext} = n_{int} + 1$ external nodes. Thus, the average external path length of an external node can be calculated as 
\begin{align}
	\overline{\mathtt{EPL}}_{0} = \frac{ \overline{\mathtt{EPL}} }{ n_{ext} } = \frac{ \overline{\mathtt{IPL}} }{n_{ext}} + \frac{ 2n_{int} }{n_{ext}}
	= 2 \sum_{k=1}^{n_{int}} \frac{1}{k} + \frac{ 2n_{int} }{n_{int}+1}.
\end{align}
Herein, $\overline{\mathtt{EPL}}_{0}$ is a function of $n_{int}$ (or equivalently, $n_{ext}$). 
{Furthermore,} $\overline{\mathtt{EPL}}_{0}$ can be viewed as the depth per an external node and the depth is averaged over all binary search trees. 

\begin{figure}[!t]
\centerline{\includegraphics[width=1\linewidth]{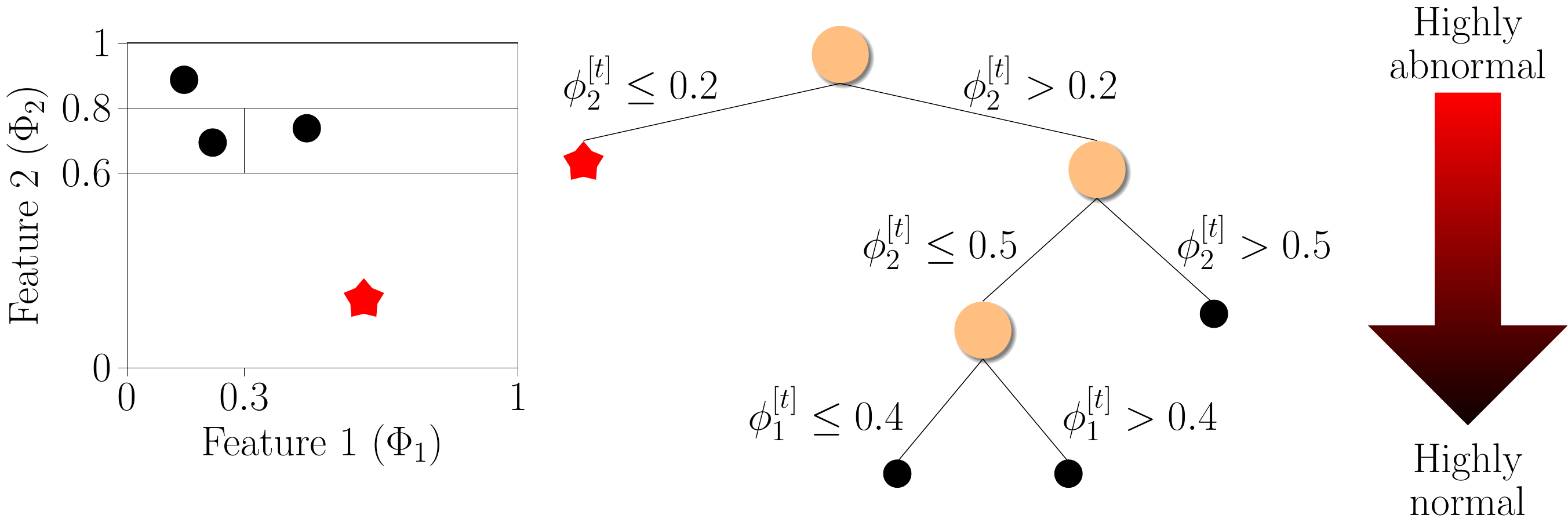}}	
	\caption{Given four instances $\{ \B{x}_{\textrm{input}}^{[t]} \}_{t=1}^4$, the isolation forest algorithm can \emph{isolate} them in four external nodes (or leaf nodes) of an isolation tree. Multiple isolation trees will then be averaged to become a single tree called isolation forest. The depth of isolation forest determines whether an instance is considered as abnormal or normal. }
	\label{fig: Forest 1}
\end{figure}
Using the above idea of binary search trees, the {authors of} \cite{Isolation_Forest} suggest building the so-called isolation trees, which then form an isolation tree. Yet another aspect is that the isolation forest algorithm {exploits} the idea of the unsuccessful search {events} in a binary search tree in order to define the path length $h(\B{x})$ of a data point $\B{x}$. By definition, an unsuccessful search traverses a path from the root node to an external node (see Fig. \ref{fig: a BST} for illustration). Similarly, in the context of an isolation tree, the path length $h(\B{x})$ is defined as \emph{the number of edges} that $\B{x}$ ``traverses from the root node until the traversal is terminated''. Using sub-sampling without replacement, the original dataset is divided into $\mathtt{NoT}$ sub-datasets, each being associated with an isolation tree. Thus there are $\mathtt{NoT}$ isolation trees in total. To find abnormal data points, \cite{Isolation_Forest} defines the so-called anomaly score as follows:
\begin{align}\label{Isolation Forest: score}
	\textrm{score}(\B{x}, \psi) = 2^{-\EX{h(\B{x})} \big/\bigr. c(\psi) } ,
\end{align}
where $\psi$ is the sub-sampling size, $c(\psi) = \left. \overline{ \mathtt{EPL}}_{0} \right|_{ n_{int} = \psi }$ is a constant, $\EX{h(\B{x})}$ is the average path length of $h(\B{x})$ and it is averaged over all $\mathtt{NoT}$ isolation trees. After calculating the score of each data point, we sort all the scores in descending order. {Given} the sorted array of scores, we can show that the first top scores correspond to anomalies. 

\paragraph{Hierarchical Clustering}
{In contrast to} $K$-means clustering, which has to predetermine the number of clusters/groups, hierarchical clustering does not require that number in advance. Instead, the number of clusters can be determined after a hierarchical dendrogram has been formed. There are two common types of hierarchical clustering, namely agglomerative and divisive hierarchical clustering:
\begin{itemize}
	\item In principle, the agglomerative approach is a bottom-up approach, which considers each data point as a separate cluster and then merges two clusters at each step until all clusters are unified as a single one. 
	\item By contrast, the divisive approach is performed in a top-down manner, where the whole data is considered as a single cluster at the beginning, and then divides the data into smaller clusters until each data point {becomes} a cluster of its own. 
\end{itemize}

\begin{figure}[!t]
	\centerline{\includegraphics[width=1\linewidth]{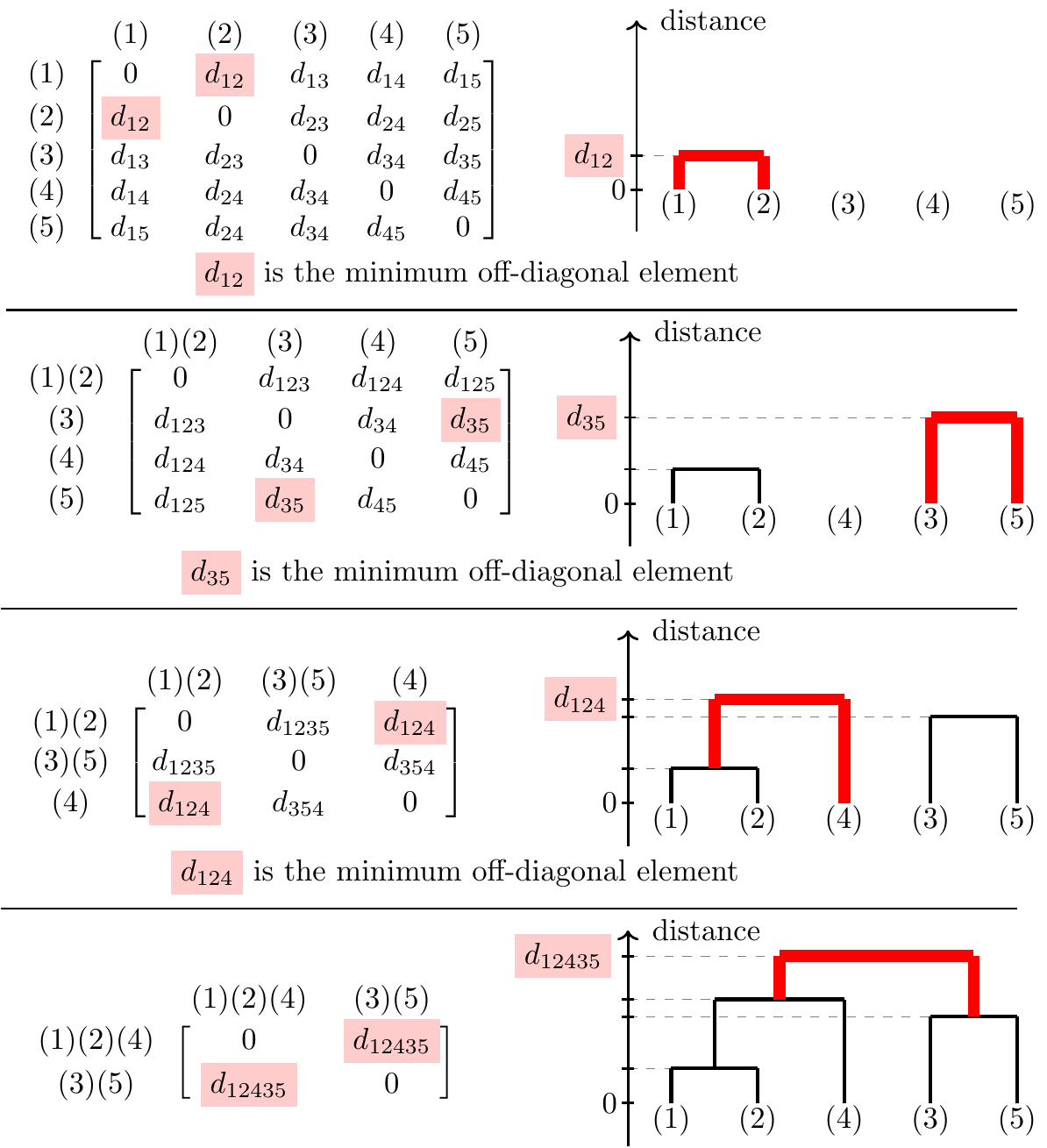}}
	\caption{A $4$-step illustration of the agglomerative hierarchical clustering. ($t$) represents $\mathbf{x}_{\textrm{input}}^{[t]}$ for $t\in\{1,2,3,4,5\}$. In each step, we find the minimum off-diagonal element in the proximity matrix and group related examples.}
	\label{fig: hierachy}
\end{figure}

Figure \ref{fig: hierachy} presents four steps in the process of constructing a simple dendrogram on the basis of the \emph{agglomerative} hierachical clustering. 
At each step, we calculate (or update) a proximity matrix and use this new matrix to decide, which two clusters will be merged. A proximity matrix is also known as a distance matrix that is symmetric.
Each off-diagonal element in a proximity matrix represents the \emph{dissimilarity} between two distinct clusters. There are many different measures of dissimilarity, such as {the} Euclidean distance, squared Euclidean distance, and Manhattan distance, just to name a few. For example, if $d_{ij}$ denotes the element at the $i$-th row, the $j$-th column of a proximity matrix, then $d_{ij}$ can be formulated as $d_{ij} = \| \B{x}_{\textrm{input}}^{[i]} - \B{x}_{\textrm{input}}^{[j]} \| $ in the case of Euclidean distance, or $d_{ij} = \| \B{x}_{\textrm{input}}^{[i]} - \B{x}_{\textrm{input}}^{[j]} \|^2 $ in the case of squared Euclidean distance. In short, the agglomerative approach uses distance measures to quantify {to what degree a} group/cluster/example is dissimilar to another. Note that distance measures are also applied to the divisive approach. 

Given a proximity matrix at each step, it is necessary to have a criterion for deciding which two clusters will be merged. There are many different criteria, which are also known as \emph{linkage} criteria. In {so-called} single linkage (or nearest neighbour) clustering, we aim to find two clusters that are closest to each other and then merge them into a new cluster. Figure \ref{fig: hierachy} is an example of single linkage in which we find the minimum off-diagonal element in a proximity matrix at each step. The corresponding row and column of that element will {identify a pair of} two clusters needed to be merged. 

In contrast to single linkage, complete linkage (or farthest neighbour) considers the maximum distance, which is determined by the most distant nodes in two clusters. It implies that clusters in complete linkage clustering tend to be smaller than clusters in single linkage clustering. {Furthermore}, there are other linkage criteria, such as the average linkage \cite{charikar2019hierarchical} and Ward linkage \cite{miyamoto2015ward}.




\subsubsection{Neural Networks} \label{SEC: NNs - CLASSIFICATION}
\begin{figure*}[!htb]
	\centerline{\includegraphics[width=.8\linewidth]{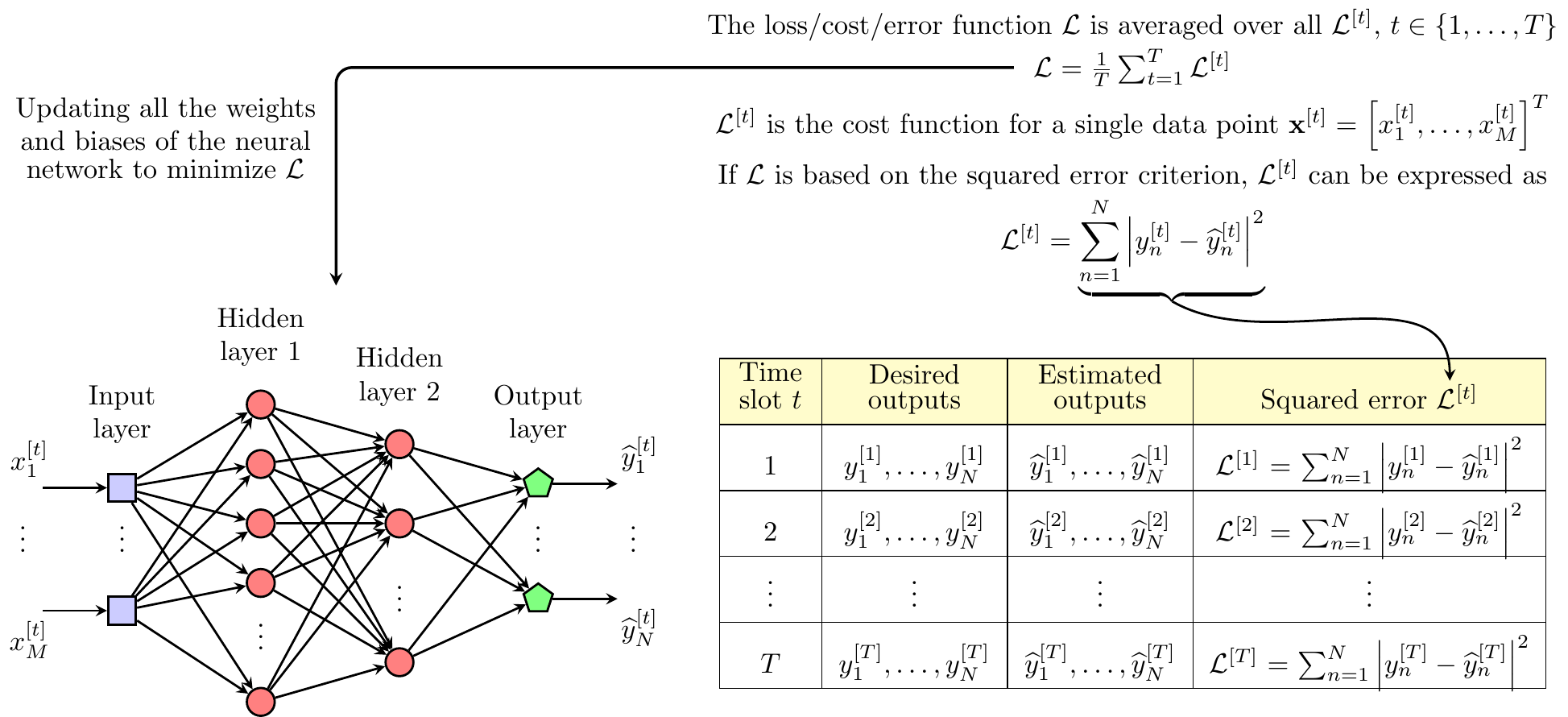}}
	\caption{A neural network with two hidden layers.}
	\label{fig: basic MLP}
\end{figure*}
Figure \ref{fig: basic MLP} describes the basic architecture of a multiple layer perceptron (MLP) {having} three layers (including an input layer, a hidden layer and an output layer). Each layer has at least one neuron, which can connect all the neurons of the previous layer to all the neurons of the next layer. 
{A detailed} investigation of the related mathematical expressions that describe the relationship between the input and output of a neuron and the relationships among neurons {can be found for example in} \cite{Goodfellow-et-al-2016}. The relationship between the input and output of the whole neural network can be described by a function {of the form} $\widehat{y}_{n} = f_{n} ( x_1, \ldots, x_M | \mathcal{W}, \mathcal{B} )$, where $x_1, \ldots, x_M$ are variables that constitute the input of a neural network, $\mathcal{W}$ is the set of link weights, and $\mathcal{B}$ is the set of biases at the neurons. The goal of the network is to find $\mathcal{W}$ and $\mathcal{B}$ {ensuring} that the functions $\{ f_{n} ( x_1, \ldots, x_M | \mathcal{W}, \mathcal{B} ) \}_{n=1}^N$ can approximate and generalize the relationship between $\{x_1, \ldots, x_M\}$ and $\{y_1, \ldots, y_N\}$, as long as the difference between the network output (i.e.,  $\{ \widehat{y}_{1}, \ldots, \widehat{y}_{N} \}$) and the \emph{actual} data (i.e., $\{y_1, \ldots, y_N\}$) is within some tolerance. This tolerance difference is {quantified} by some loss function (known as error/cost/objective function).

Since the loss function {quantifies} how well a neural network works, it will be calculated at the output of the network. There are various types of cost functions each resulting in a different performance \cite{Janocha2017, Golik2013, Kline2005}. A pair of well-known cost functions are {constituted by the} squared-error and cross-entropy functions. Depending on the assessment criteria (such as a posterior probability estimation or convergence), different types of cost functions may be preferred over others. In general, the goal of a neural netwrok is to minimize the loss function by {appropriately} adjusting the network parameters. This will often lead to solving a minimization problem (whose objective function is the loss function) with respect to the network parameters (note that these parameters are in the sets $\mathcal{W}$ and $\mathcal{B}$ as mentioned above). Since the network parameters stick with the operation of neurons, the choice of activation functions for each and every neuron {deserves special} consideration. 

In general, a multiple layer perceptron {relies on} neurons each being activated by a certain activation function $\sigma(\cdot)$. In fact, the activation functions used in a neural-network-based classifier have direct effects on the classification performance \cite{Shenouda2006, Pedamonti2018}. {Similarly to the diverse variety of} loss functions, there are many types of activation functions.
In a neural network, it is theoretically possible to combine different types of activation functions, including but not limited to a binary step, linear, sigmoid, tanh and ReLU functions.
Among them, ReLU is one of the most widely-used activation functions \cite{Lecun2015}. {Furthermore}, there is also a huge range of other activation functions, {such as the} maxout \cite{Goodfellow2013}, softmax \cite{LiangXuezhiandWang2017}, piecewise linear \cite{Agostinelli}, and distance-based activation functions \cite{Duch-survey-activation_functions}, just to name a few. 

As for the input layer, the number of neurons should be equal to the number of features {in the} data input. For example, we may wish to create a multiple layer perceptron, which allows a column vector $\textbf{x}_{\textrm{input}}^{[t]}$ to pass through. Then, the number of input neurons should be equal to the length of $\textbf{x}_{\textrm{input}}^{[t]}$. Recall that $\textbf{x}_{\textrm{input}}^{[t]} = [ x_1^{[t]}, \ldots, x_M^{[t]} ]^T$ is the $t$-th data point, and each element in $\textbf{x}_{\textrm{input}}^{[t]}$ represents a feature (see Section \ref{SEC: DATA} for more details). Regarding the output layer, {a single} neuron may be {capable performing} a binary classification task. For example, if the sigmoid function is applied to the unique neuron of the output layer, we {may infer} that the network is under attack when the value of the output is {higher} than $1/2$; otherwise, the network is still safe. In the case of more {complex} multi-class classification, {using a single} neuron at the output layer may not be {sufficient for a confident decision. Indeed, it is plausible that} multiple output neurons are necessary {for making multi-class decision. Furthermore,} other types of activation function should be applied, such as the softmax function \cite{LiangXuezhiandWang2017, Dunne1997}. 



\subsubsection{Reinforcement Learning} \label{Reinforcement}
Reinforcement learning enables a system to interact with the environment and learn from these interactions. Normally, reinforcement learning problems can be formalized by the framework of a Markov decision process \cite{Pareto_Wang2020, Luong_RL_2019}.\footnote{Markov decision processes are used for decision making.} Intuitively, reinforcement learning considers a certain agent that learns a policy. The agent first observes its current state and then it takes an action. Based on the action to be taken, the environment gives the agent some feedback {based on} the new state and the reward. Using the feedback obtained, the agent will take {its} next action. This process is iteratively executed many times in order to find the optimal policy that maximizes the expected total reward in the long run. 

{Some of the} popular reinforcement learning algorithms are Q-learning \cite{watkins1992q}, SARSA \cite{Alfakih-Access2020-SARSA} and deep Q-learning \cite{fan2020theoretical}. While Q-learning {operates} in an offline fashion that attains the optimal policy after the whole algorithm converges, SARSA is an online learning-based algorithm that can find the optimal action at each {individual} iteration \cite{Luong_RL_2019}. {Hence,} Q-learning may be {more} suitable for a small space of states and actions. On the other hand, deep Q-learning, which {relies on} the combination of reinforcement learning and neural networks, {relies on} a deep Q-network {for solving} high-dimensional problems \cite{Mnih2015}. 

\subsubsection{Other Advanced ML Algorithms Learning} \label{Other ML algorithms}
\paragraph{Variational Bayesian Learning}
Let $\B{x}$ represent a data sample and $\B{z}$ a vector of parameters. Using Bayes' theorem, we have the following {\em aposteriori} probability:
\begin{align}
    \underbrace{
    p(\B{z} | \B{x}) 
    }_{\text{posterior}}
    = 
    \underbrace{
    p(\B{x}|\B{z}) 
    }_{\text{likelihood}}
    \times
    \underbrace{
    p(\B{z}) 
    }_{\text{prior}}
    / 
    \underbrace{
    p(\B{x})
    }_{\text{evidence}} .
\end{align}
{If $\B{z}$ is continuous, then the denominator $p(\B{x})$ can be calculated as
$
    p(\B{x}) = \int_{\B{z}} p(\B{x}, \B{z}) d\B{z}.
$

According to Jensen's inequality, we have $\log\left(\EXs{p(\B{x})}{f(\B{x})}\right) \geq \EXs{p(\B{x})}{ \log\left( f(\B{x}) \right) }$. On the other hand, $\log \left(p(\B{x})\right)$ can be written as the logarithm of an expectation, i.e.
\begin{align}
    \log \left(p(\B{x})\right) 
    &= \log \left( \int_{\B{z}} p(\B{x}, \B{z}) d\B{z} \right)
    = \log \left( \int_{\B{z}} Q(\B{z}) \frac{p(\B{x}, \B{z})}{Q(\B{z})} d\B{z} \right)
    \nonumber \\ &
    = \log \left( \EXs{Q(\B{z})}{ \frac{p(\B{x}, \B{z})}{Q(\B{z})} } \right) .
\end{align}
Thus, applying Jensen's inequality to $\log \left(p(\B{x})\right)$, we arrive at:
\begin{align}
    \underbrace{
    \log \left(p(\B{x})\right) 
    }_{\text{evidence}}
    \geq
    \underbrace{
    \EXs{Q(\B{z})}{ 
    \log \left(
    \frac{p(\B{x}, \B{z})}{Q(\B{z})} 
    \right)
    } 
    }_{\text{evidence lower bound (ELBO)}}
    \triangleq L(\B{z}).
\end{align}
More importantly, the difference between the log marginal probability of $\B{x}$ and the ELBO turns out to be the Kullback-Leibler (KL) divergence \cite{fox2012tutorial} of
\begin{align}
    \log \left(p(\B{x})\right) - L(\B{z}) 
    = \text{KL} \left[ Q(\B{z}) || p(\B{z} | \B{x}) \right] .
\end{align}
The KL divergence is widely used for quantifying how much a probability distribution differs from another one. Hence $\text{KL} \left[ Q(\B{z}) || p(\B{z} | \B{x}) \right]$ quantifies the difference/dissimilarity, between the \emph{variational} probability distribution $Q(\B{z})$ and the \emph{true aposteriori probability} $p(\B{z} | \B{x})$. In this context maximizing the ELBO is equivalent to minimizing the KL divergence. The core idea here is to find some tractable distribution $Q(\B{z})$ (e.g. the Gaussian distribution) that approximates the {\em aposteriori} distribution $p(\B{z} | \B{x})$. Based on this, sophisticated variational Bayes (VB) ML algorithms have been proposed, which can be categorized into mean-field VB (MFVB) and fixed-form VB (FFVB)~\cite{tran2021practical}.

Some of the most popular VB ML algorithms are constituted by the family of \textit{variational auto-encoders} (VAEs) proposed in the pioneering contribution~\cite{kingma2013auto}. A typical VAE includes an encoder that yields an approximate {\em aposteriori} distribution $p_{\boldsymbol\theta}(\B{z}|\B{x})$, and a decoder that yields a likelihood distribution $p_{\boldsymbol\phi}(\B{x}|\B{z})$. Herein, $\boldsymbol\theta$ denotes all the weights of the encoder, while $\boldsymbol\phi$ represents all the weights of the decoder. By training the VAE, the parameters $\boldsymbol\theta$ and $\boldsymbol\phi$ are optimized so that the approximate {\em aposteriori} distribution $p_{\boldsymbol\theta}(\B{z}|\B{x})$ becomes similar to the true {\em aposteriori} distribution.




\paragraph{Mixture of experts}
When the data volume becomes large, it is necessary for an AI-aided system to be able to scale up its training models. One of the popular methods of scaling up AI models is the so-called mixture of experts (MoEs) technique~\cite{MoE_1, MoE_2}. Briefly, the MoE method is a neural-network-based ensemble learning technique. Similar to other ensemble methods, the MoE is based on the 'divide-and-conquer' principle. Herein, the underlying idea behind the MoE is that the global input space is divided into smaller sub-spaces each being assigned to a specific local model for training. Then all the outputs of the local models will be combined. Note that each local model is termed as an expert. Furthermore, the MoE consists of a neural network, referred to as the gating network, that supervises the 'divide-and-conquer' process.

Figure \ref{fig: MoE illustration 1} illustrates the partitioning of a dataset into two smaller datasets that will then be trained by a pair of local experts in support of the classification task. This results in beneficial specialization, because each local model (i.e., each expert) will specialize in different small tasks by training and learning based on different small input spaces. Based on the outputs of the experts, a gating network will be trained based on the global input data. Herein, the role of each expert is to learn from a local input space, while the role of the gating network is to learn from the outputs of experts. Indeed, the MoE functions in a 'divide-and-conqure' manner so that the input space may be beneficially divided into subsets - each being assigned to a suitably specialized expert for training.

\begin{figure}[!htb]
	\centerline{\includegraphics[width=1\linewidth]{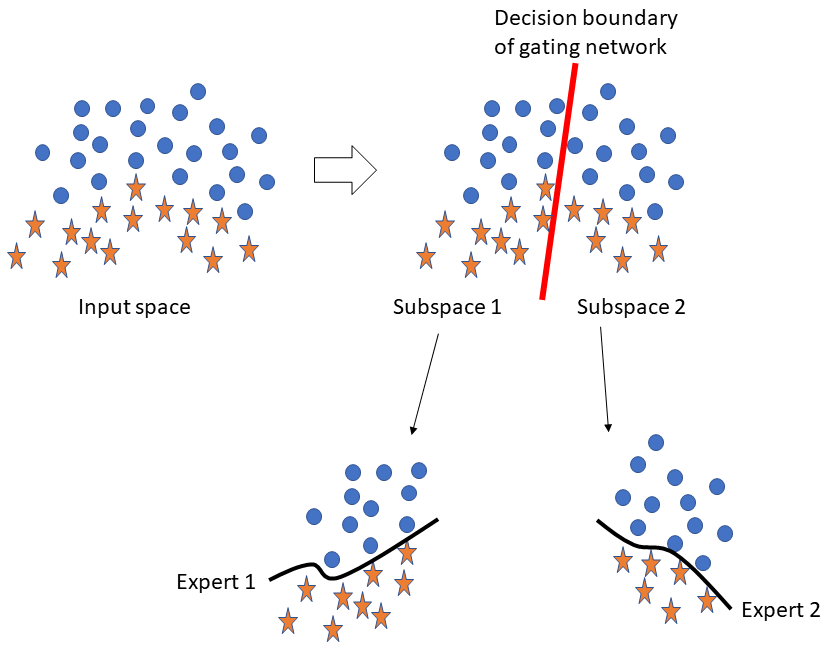}}
	\caption{An illustration of using MoE for classification.}
	\label{fig: MoE illustration 1}
\end{figure}

\paragraph{Meta-learning}
When it comes to learning from prior experience in the context of ML, meta-learning emerges as a compelling candidate, since it represents the concept of ``\textit{learning to learn}''~\cite{meta-learning-1, meta-learning-2}. Naturally, we do not always have to learn new skills from scratch once we have suceeded in learning some other skills. In other words, after learning or inferring a specific skill from a task, it becomes easier to learn another skill based on a new task, because we ``\textit{learn how to learn}'' by relying on skills transferred across tasks~\cite{meta-learning-2}.

To elaborate a little further by describing the concept of meta-learning mathematically, let us consider a set of previous tasks $\{t_1, t_2, \ldots, t_M\}$, a set of parameter vectors $\{\boldsymbol\theta_1, \boldsymbol\theta_2, \ldots, \boldsymbol\theta_N\}$, and a set of system performance evaluations $E_{\text{previous}} = \{e_{11}, e_{12}, \ldots, e_{mn}, \ldots, e_{MN}\}$ with $m\in\{1, \ldots, M\}$, $n\in\{1, \ldots, N\}$. Herein, a task $t_m$ can be also be viewed as a learning algorithm, e.g., a neural network based learning processed harnessed for optimizing some function. In this context each parameter vector $\boldsymbol\theta_n$ represents a specific configuration of an ML model, e.g., it represents all the weights of a neural network. Finally, each value $e_{mn}$ is a value of the function $e(t, \boldsymbol\theta)$ evaluated at $t=t_m$ and $\boldsymbol\theta = \boldsymbol\theta_n$, e.g., $e_{mn}$ can be the accuracy of an ML system. Let us assume now that we want to predict a new parameter vector $\boldsymbol\theta_{\text{next}}$ for the new task $t_{\text{next}}$, given that the new set of performance evaluations - namely $E_{\text{next}} = \{ e_{\text{iter}, 1}, e_{\text{iter}, 2}, \ldots \}$ - can be obtained in an iterative manner during the training. Based on the concept of meta learning, we will train a new ML model based on the meta data $E_{\text{previous}} \cup E_{\text{next}}$. For further details, please refer to~\cite{meta-learning-1, meta-learning-2}, for example.

\paragraph{Self-supervised learning (SSL)}
In contrast to supervised learning that relies on labelled data, self-supervised learning (SSL) does not require the data to be labelled and thus it is normally referred to as a branch of unsupervised learning. However, while most of the popular unsupervised learning algorithms aim for detecting the pattern of data and clustering datasets, SSL is centered on recovering and self-generating the data~\cite{SSL-2023-Xiao}. It may also be observed that SSL algorithms tend to be constructed based on the architecture of neural networks.

To elaborate a little further, the architecture of SSL may be divided into two main categories: i) Generative SSLs and ii) Contrastive SSLs. As for generative SSLs, an encoder is used for encoding an input sample $x$ iinto $z$, and then a decoder will process $z$ to reconstruct $x$. A typical model using generative SSLs is constituted by the auto-encoder (AE)~\cite{tschannen2018recent}. On the other hand, as for contrastive SSLs, after decoding $x$ into $z$, the similarity/dissimilarity will be quantified for differentiating the input sample from others~\cite{jaiswal2020survey}. According to~\cite{SSL-2023-Xiao}, there are also hybrids of generative and contrastive solutions, including the family of generative adversarial networks (GANs). Following their introduction in 2014~\cite{goodfellow2020generative}, GANs have received substantial attention as a benefit of their applicability in numerous fields, such as image processing and natural language processing. In the architecture of a GAN, a generator and a discriminator are simultaneously trained in a competing manner~\cite{GANs-an-overview-2018}. While the generator is trained for generating new samples, the discriminator is trained for differentiating the actual sample and the generated (fake) samples. Herein, the goal of a GAN is to confuse the discriminator in distinguishing the actual samples from fake ones and to allow the generator to create fake-but-plausible data that even humans find hard to realize.

\paragraph{Explainable AI}
Despite the success of recent ML algorithms, there are still concerns about applying ML in a variety tasks, such as decision-making and recommender systems. Sometimes it is hard to interpret the results returned by an ML algorithm due to the lack of transparency and reasoning mechanisms. Indeed, ML models like neural networks are commonly viewed as ``black box'' models, mainly because their results are not always explainable by humans~\cite{XAI-Hagras-2018}. Thus, explainable AI (XAI) is expected to be one of the next steps in the realms of AI development, where human users can understand more about the results returned by ML models.

There are diverse criteria for declaring an AI solution to be explainable, including transparency, causality and trust. Herein, transparency implies the capability of explaining the ML models and their results to both technical and non-technical users. Increasing the transparency of an ML model will help even non-technical users to interpret how the model works and what to expect upon using it. By contrast, causality implies that causal inferences may be drawn from the behaviour of ML models. Finally, trust quntifies to what degree human users can believe in a certain ML model. In general, explainable AI aims for improving the explainability in order to provide users with an improved understanding and to lend the users confidence when applying the ML models~\cite{XAI-Hagras-2018, XAI-lundberg2020local}.}

\subsection{Recent Advances and Future Directions}
We divide this sub-section into two parts corresponding to the two main aspects of ML-aided PHY security: i) ML-aided PHY authentication and ii) ML-aided PHY security design. For each part, we {critically appraise the literature of} ML-aided PHY authentication (or ML-aided PHY security design) and then discuss future directions. 
\subsubsection{\textbf{ML-aided PHY Authentication} }

\paragraph{Recent Advances} 

\begin{table}[!t]
	\centering
	\caption{The Application of ML Algorithms in PHY Authentication}
	\renewcommand{\arraystretch}{1.5}
	\begin{tabular}{|c|c|}		
\hline 
\multicolumn{1}{|c|}{ \multirow{2}{*}{ Papers } }
& \multicolumn{1}{c|}{ \multirow{2}{*}{Algorithms} }
\\  
{  } 
& {  } 
\\ \hline
\cite{PHY-Authentication_Spoofing-attacks_2018} \cite{PHY_authentication_7_people} 
& $k$-NN  
\\ \hline
\cite{PHY-Authentication_Spoofing-attacks_2018} \cite{Hoang2020_WCL} 
& $k$-means 
\\ \hline
\cite{Hoang2020_WCL} 
& OC-SVM
\\ \hline
\cite{PHY-Authentication_ML_1} \cite{PHY_authentication_7_people} 
& SVM
\\ \hline
\cite{PHY-Authentication_ML_1} 
& LDA
\\ \hline 
\cite{PHY_authentication_7_people} 
& Decision \& Bagged Trees 
\\ \hline 
\cite{PHY-Authentication_ML_4} 
& One-Class Nearest Neighbour
\\ \hline 
\cite{PHY_authentication_Logistic_Regression} 
& Logistic Regression 
\\ \hline 
\cite{PHY-Authentication_ML_3} \cite{PHY-Authentication_ML_GaussianMixture_2018} \cite{PHY-Authentication_KLT_2019} 
& Gaussian Mixture Model 
\\ \hline 
\cite{PHY_authentication_Wang2017}, \cite{PHY-Authentication_Data-Augmentation}, \cite{PHY_authentication_Manufactoring_Process_2019} 
& Neural Network 
\\ \hline 	
\cite{PHY-Authentication_Reinforcement} 
& Reinforcement Learning  
\\ \hline
\cite{PLA_Lajos_Fang2019} 
& Kernel-based method 
\\ \hline 
\end{tabular}
\label{table: ML PHY authentication}
\end{table}
A range of ML-aided classification algorithms {may} also be applied {for enhancing} security at the physical layer. For example, Table \ref{table: ML PHY authentication} lists the {popular} ML algorithms used and the related papers. In the following, we review these papers in more detail. 

In \cite{PHY-Authentication_Spoofing-attacks_2018}, a {beneficial} combination of both $k$-NN and $K$-means techniques is suggested {for detecting} spoofing attacks in wireless sensor networks. $K$-means clustering is used for extracting features from RSS samples, while $k$-NN plays the role of a classifier \cite{abu2019effects}. 
{The authors of} \cite{Hoang2020_WCL} compare the performance of $K$-means and OC-SVM, and show that OC-SVM results in a more stable detection performance than $K$-means, when the power of a spoofing signal {fluctuates}. However, {the performance of} OC-SVM is worse than {that of} $K$-means, when the spoofing signal is transmitted at high power. In \cite{PHY-Authentication_ML_1}, both SVM and LDA are {harnessed for processing} the RSS, TOA and another correlation-based feature. 
{As a further development, the authors of} \cite{PHY_authentication_7_people} suggest using estimated channel matrices for generating features and then compare four different ML algorithms, i.e., the $k$-NN, SVM, decision tree and bagged tree, in terms of {their} accuracy and prediction time.
Similar to \cite{Hoang2020_WCL}, the training data considered in \cite{PHY-Authentication_ML_4} contains only {a single} class under the assumption that there is no knowledge {concerning} any of the potential eavesdropper. In \cite{PHY-Authentication_ML_4}, single-class nearest neighbour classification is performed on the single-class data {for finding both} high- and low-density regions, thereby creating a predictive model for authentication. {The authors of} \cite{PHY_authentication_Logistic_Regression} propose a logistic regression model for authentication and estimate the coefficients of the model by using the popular Frank–Wolfe algorithm.

{Furthermore, the} Gaussian mixture model is proposed in \cite{PHY-Authentication_ML_3, PHY-Authentication_ML_GaussianMixture_2018, PHY-Authentication_KLT_2019}. To elaborate, Weinand \textit{et al.} \cite{PHY-Authentication_ML_3} build a simple authentication framework {for} machine type communication by using the magnitude of channels as the input data. Qiu \textit{et al.} \cite{PHY-Authentication_ML_GaussianMixture_2018} build the training data {associated} with two features that are based on the Euclidean distance and Pearson correlation, respectively. Qiu \textit{et al.} \cite {PHY-Authentication_KLT_2019} exploit the Karhunen-Loeve transform {for extracting} the most significant
features and {for} reducing the dimension of the data. The low-dimensional channel representation created by the Karhunen-Loeve transform shows the superiority of the framework in \cite{PHY-Authentication_KLT_2019} over the similar framework of \cite{PHY-Authentication_ML_3}. The spoofing detection performance comparison between \cite{PHY-Authentication_KLT_2019} and \cite{PHY-Authentication_ML_3} once again confirms that {bespoke} feature selection is always crucial in the data preprocessing {used} for authentication. 

{The family of} neural networks is considered in \cite{PHY_authentication_Wang2017, PHY-Authentication_Data-Augmentation, PHY_authentication_Manufactoring_Process_2019}. Specifically, in \cite{PHY_authentication_Wang2017}, the data is formulated in a similar way to \cite{PHY-Authentication_ML_GaussianMixture_2018}, {apart from the} difference that {Wang} \textit{et al.} \cite{PHY_authentication_Wang2017} use extreme machine learning based on neural networks, while {Qiu} \textit{et al.} \cite{PHY-Authentication_ML_GaussianMixture_2018} use a Gaussian mixture model. {Liao} \textit{et al.} \cite{PHY-Authentication_Data-Augmentation} focus {their attention} on data augmentation methods that {harness} an extra amount of data in addition to the existing data. The data augmentation {is eminently suitable for} improving the robustness of the data, especially when {dealing with high-volume} data. {Chatterjee} \textit{et al.} \cite{PHY_authentication_Manufactoring_Process_2019} consider the identification of nodes in an Internet-of-Things system by developing a detection framework based on neural networks and physical unclonable functions. Regarding reinforcement learning, {Xiao} \textit{et al.} \cite{PHY-Authentication_Reinforcement} model the interaction between a legitimate user and an eavesdropper as a zero-sum authentication game in a dynamic environment, and {employ} reinforcement learning for finding the optimal threshold of hypothesis testing. Finally, {Fang} \textit{et al.} \cite{PLA_Lajos_Fang2019} develop an adaptive learning model based on the kernel least mean square method in which the authentication problem is modelled as a linear system {for reducing} the dimension of the feature space and complexity.  


\paragraph{Future Directions}
In general, there is no {consensus in the literature concerning the best} criteria for creating the input data in previous works.
For instance, the authors of \cite{PHY_authentication_Wang2017} and \cite{PHY-Authentication_ML_GaussianMixture_2018} create the data based on the Euclidean distance and the Pearson correlation, respectively. {By contrast,} the data for the PHY authentication in \cite{PHY_authentication_Logistic_Regression} relies on the RSSI. 
The features of the data used for PHY authentication in \cite{PHY_authentication_Manufactoring_Process_2019} include the local oscillator frequency, Doppler shift, {as well as} the in-phase and quadrature components of transmitted signals. 
{By contrast, the} carrier frequency offset, channel impulse response, and RSSI are {used} as features for ML-aided PHY authentication in \cite{PLA_Lajos_Fang2019}. Given that the selection (or extraction) of features for the data directly affects on the detection performance of an ML-aided PHY authentication model as shown in \cite{PHY-Authentication_KLT_2019} and \cite{PHY-Authentication_Data-Augmentation}, future works should pay more attention to the topic of feature selection. 

{However, at the time of writing there are no authoritative} comparison of the security performance of ML-aided classification algorithms. Thus future {research has} to carry out extensive studies to find suitable input data types for PHY authentication, and to find suitable ML classification algorithms for each type of data. 
{On a similar note}, anomaly detection methods {require similar attention to that concerning} PHY security. As part of the ML {family}, anomaly detection (also known as outlier detection) includes many methods that are developed for identifying rare events/observations and thus they are normally used for intrusion detection {upper ISO} layers \cite{Cyber_ML_Buczak2016}. However, anomaly detection methods remain underused in {maintaining} security at the physical layer. Additionally, {given} the {ongoing} development of neural networks, it is shown that neural networks can be {readily} exploited {for discovering} hidden-but-useful features from available data for authentication purposes \cite{Survey_Deep_Mao2018, Survey_Deep_Chen2019}. Finally, it is also worth considering the {class of} \emph{cross-layer} authentication {harnessing} the data {both at the physical and upper layers.}

\subsubsection{\textbf{ML-aided PHY Security Design}}
\paragraph{Recent Advances}
When it comes to the integration of ML into ML-aided PHY security designs, {a representative range of contributions are} listed in Table \ref{table: ML PHY security design}, {which are detailed} below. 


\begin{table}[!t]
	\centering
	\caption{The Application of ML Algorithms in PHY Security Design}
	\renewcommand{\arraystretch}{1.5}
	\begin{tabular}{|c|c|}		
		\hline 
		\multicolumn{1}{|c|}{ \multirow{2}{*}{ Papers } }
		& \multicolumn{1}{c|}{ \multirow{2}{*}{Algorithms} } 
		\\  
		{  } 
		& {  } 
		\\ \hline
		\cite{He2019-PLS-ML} \cite{PLS_Information-theoretic_Besser2019} \cite{PLA_Xing2019} 
		&  Neural Network 
		\\ \hline 
		\cite{PLS_Reinforcement_Li2019} \cite{ML-PHY-Design_Reinforcement_Xiao2019} \cite{PLS_Reinforcement_Miao2019} 
		& Reinforcement Learning 
		\\ \hline 
		\cite{SVM_He2018} 
		& SVM, naive Bayes
		\\ \hline 	
		\cite{ML-PHY_Design_Wen2019} 
		& \emph{not to be named}  
		\\ \hline 
	\end{tabular}
	\label{table: ML PHY security design}
\end{table}

He \textit{et al.} \cite{He2019-PLS-ML} exploit both beamforming and artificial noise to deal with an eavesdropper and formulate an optimization problem that maximizes the effective secrecy throughput {by harnessing} a neural network. 
Xing \textit{et al.} \cite{PLA_Xing2019} also use neural networks for secure transmission {by relying on} cooperative beamforming in a relay-aided system. 
Besser \textit{et al.} \cite{PLS_Information-theoretic_Besser2019} strike a trade-off between reliability and security by {formulating a} multiple-objective optimization problem. To {resolve the associated} trade-off, wiretap codes are designed based on neural network-aided autoencoders.
{As another development, Li} \textit{et al.} \cite{PLS_Reinforcement_Li2019} design a Q-learning-based power control strategy for secure transmission {by considering a powerful} attacker {having a high} number of antennas. 
{Upon} using reinforcement learning, Xiao \textit{et al.} \cite{ML-PHY-Design_Reinforcement_Xiao2019} propose an optimal beamforming scheme for visible light communication
In \cite{PLS_Reinforcement_Miao2019}, reinforcement learning is used by {Miao and Wang} to handle the frequency allocation problem without {requiring any} information exchange among base stations. 
{In contrast to} the above-mentioned papers, He \textit{et al.} \cite{SVM_He2018} do not use neural networks or reinforcement learning in designing secure transmissions. Instead, an SVM and a naive Bayes algorithm are used for transmit antenna selection. By selecting the {most} suitable antenna for transmission, the security level is shown to be improved. 
By contrast, Wen \textit{et al.} \cite{ML-PHY_Design_Wen2019} assume that intelligent attackers can use supervised learning to decode {even} artificial-noise-{contaminated} signals.

\begin{figure*}[t!]	\centerline{\includegraphics[width=.75\linewidth]{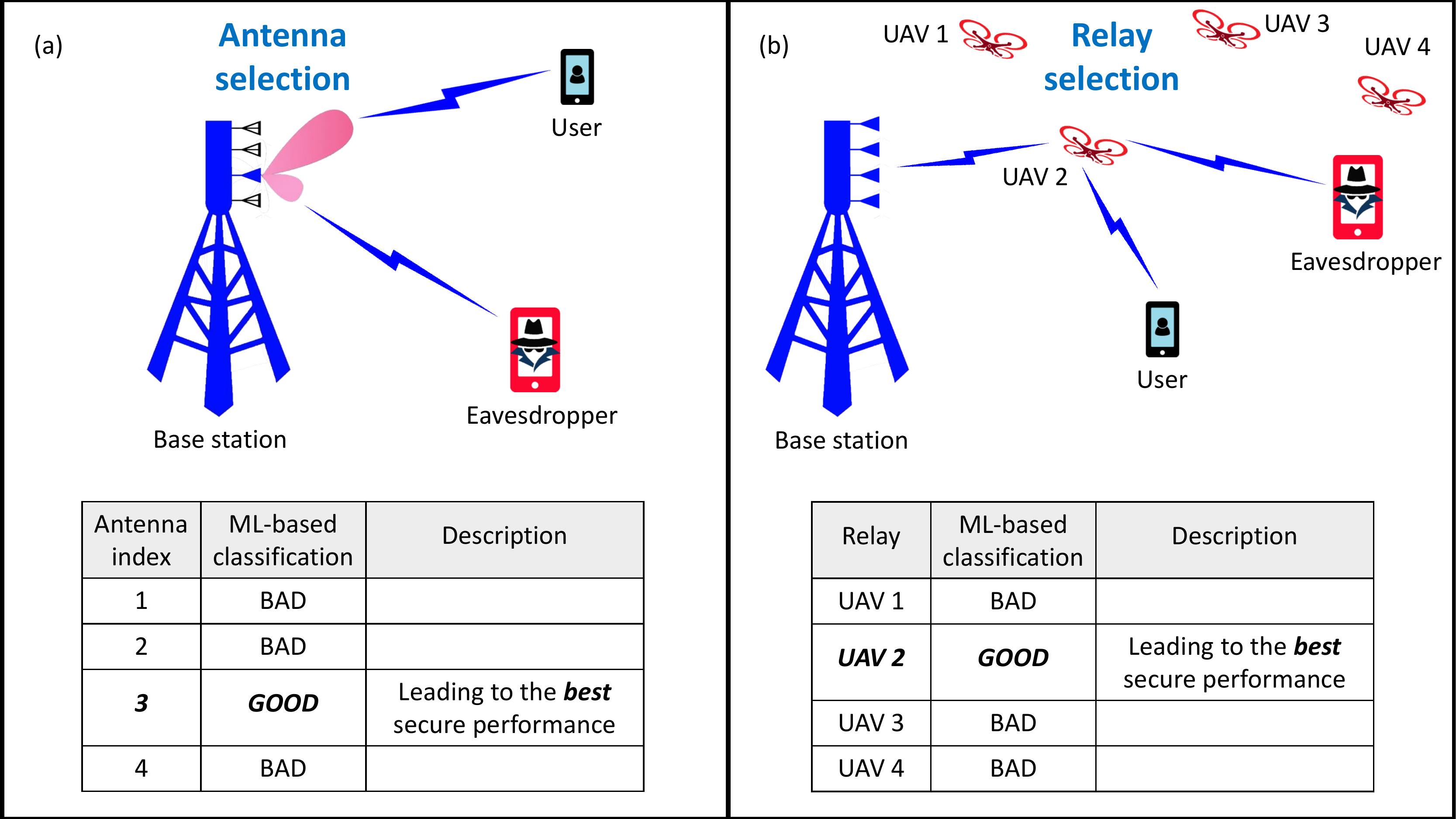}}
	\caption{Two examples of using classification algorithms (i.e., ML classifiers) for selecting the best system components. In the sub-figure (a), the result of ML-based classification shows that the $3$-rd antenna can lead to the best secure transmission. In the sub-figure (b), the result of ML-based classification shows that the $2$-nd UAV can be the best relay for retransmitting signals to the intended user.}
	\label{fig: classification for system designs}
\end{figure*}

\paragraph{Future Directions}
In terms of PHY security design, ML classification algorithms are {capable of going beyond} realizing eavesdroppers. More particularly, ML classifiers can be used for any classification tasks rather than {being limited to} eavesdropping detection. For example, based on the system component classification, we can decide to use the most suitable components {for attaining an increased} security level. To {argument this further,} let us consider the work of He \textit{et al.} \cite{SVM_He2018}. The idea of \cite{SVM_He2018} is to use SVM and naive Bayes algorithms to classify the transmit antennas of a transmitter in order to find the best antenna for secure transmission. This idea {may also be readily} generalized to other system component selection, because there are many ways of enhancing the security performance through {the selection of the most appropriate} system components, {such as} antenna selection \cite{Antenna-Selection_Security_2012}, transmitter selection \cite{Transmitter-Selection_Security_2019}, relay selection \cite{Relay-Selection_Security_2015, Security-reliability_Zou1}, and so on \cite{Survey_Security_Multi-Antenna}. Note that the {authors of} \cite{Antenna-Selection_Security_2012, Relay-Selection_Security_2015, Transmitter-Selection_Security_2019} do not consider the {employment} of ML algorithms for component selection. To visualize the generalized idea of using ML for selecting system components, let us consider Figure \ref{fig: classification for system designs}, {where a pair of} security designs are considered: (a) using ML classifiers for selecting the best transmit antenna, and (b) using ML classifiers for selecting the best relay. In short, the use of ML classifiers can point out which system components are the best for secure and reliable transmission. {However, at the time of writing}, there is {paucity of literature} in this research direction.
 
\section{PHY Security Optimization\\by Neural Networks}\label{SEC: Design - NN}
This section emphasizes the potential {benefits} of neural networks in optimizing PHY security. More specifically, we {bridge} the gap between 
\begin{itemize}
	\item \emph{complex}-variable optimization problems to be faced in PHY security (see Sub-section \ref{sub-sec: PHY security OPT})
	\item and \emph{real}-variable optimization problems to be solved by \emph{real}-valued neural networks (see Sub-section \ref{sub-sec: real-valued NNs}).
\end{itemize}
Then we discuss the ML-oriented research {issues of} \emph{complex}-variable based neural networks, thereby showing the potential benefits of \emph{complex}-valued neural networks {in} handling PHY security optimization (see Sub-section \ref{sub-sec: NNs: complex, Pareto}). Finally, multiple-objective optimization is also discussed {as a promising} future research {issue} (see Sub-section \ref{sub-sec: NNs: complex, Pareto}).

\subsection{Optimization in PHY Security Design}\label{sub-sec: PHY security OPT}
Let us consider a simple security system in which a base station (A) broadcasts its signals to a legitimate user (B), {which is overheard by} an eavesdropper (E). Let us denote the A-B channel and the A-E channel by $\B{h}_B$ and $\B{h}_E$, respectively. {Upon} denoting the \emph{instantaneous} capacity of the A-B channel and that of the A-E channel by $C_B$ and $C_E$, respectively, the \emph{instantaneous} secrecy rate can be expressed as 
\begin{align}
C_s &= \max(0, C_{\textrm{B}} - C_{\textrm{E}})
= \begin{cases}
    0, & \text{if~} C_{\textrm{B}} \leq C_{\textrm{E}}\\
    C_{\textrm{B}} - C_{\textrm{E}}, & \text{if~} C_{\textrm{B}} > C_{\textrm{E}}.
  \end{cases}.
\end{align}
Secure system designs {typically} deal with the quantity $\Delta = C_{\textrm{B}} - C_{\textrm{E}}$, which is the difference between the capacity of Bob's channel and that of Eve's channel. Based on the definitions of $\Delta$, $C_{\textrm{B}}$ and $C_{\textrm{E}}$, we will formulate optimization problems with {the associated} power constraints in mind. {Upon} denoting the transmit vector of Alice by $\B{s}$ and {bearing in mind} the transmit power limitation, each and every signal should satisfy the following constraint:
\begin{align}
	\EX{\|\B{s}\|^2} \leq P_{max} ,
\end{align}
where $P_{max}$ is the power budget of Alice. 

It is {plausible} that $C_B$ is a function of $\B{s}$ and $\B{h}_B$. At the same time, $C_E$ is a function of $\B{s}$ and $\B{h}_E$. Consequently, $\Delta$ is a function of $\B{s}$, $\B{h}_B$ and $\B{h}_E$. As such, we can either write $C_B$, $C_E$ and $\Delta$ for short in the way we have presented {them} above, or write $C_B(\B{s}, \B{h}_B)$, $C_E(\B{s}, \B{h}_E)$ and $\Delta(\B{s}, \B{h}_B, \B{h}_E)$ to indicate their dependence on both the transmit signal and the channels. Since we are unable to control the channels, the overall goal of optimization becomes {that of appropriately} designing the transmit signal $\B{s}$. Table \ref{tab: opt probs} illustrates four different optimization problems that are widely considered in the literature. The problems in Table \ref{tab: opt probs} are not necessarily convex. It should also be noted that (P2) and (P4) may be viewed as \emph{security} {vs.} \emph{reliability} trade-off problems, because $C_B$ and $C_E$ {constitute a pair of} conflicting functions. Roughly speaking, $C_B$ reflects the reliability of transmission, and $C_E$ {characterizes} the ability of an eavesdropper to decode the source signal. While increasing the transmit power improves the reliability {by reducing the bit error rate, the eavesdropper also gets a better chance of detection, which} degrades the security level \cite{Security-reliability_Zou1, Security-reliability_Zou2}. 

\begin{table}[!t]
\centering 
\caption{Types of Optimization Problems in PHY Security}
\renewcommand{\arraystretch}{1.5}
\begin{tabularx}{.7\linewidth}{|ccX|} 
\hline 
\multicolumn{3}{|c|}{\textbf{MAXIMIZATION}}
\\ \hline 
(P1) &$\underset{\B{s}}{\text{max}}$ &$\Delta(\B{s},\B{h}_B,\B{h}_E)$ 
\\
{} &s.t. &$\EX{\|\B{s}\|^2} \leq P_{max}$  
\\ \hline
(P2) &$\underset{\B{s}}{\text{max}}$ &$C_{\textrm{B}}(\B{s}, \B{h}_B)$  
\\
{} &s.t. &$C_{\textrm{E}}(\B{s}, \B{h}_E)\leq$ a threshold 
\\
{} &{} &$\EX{\|\B{s}\|^2} \leq P_{max}$ 
\\ \hline \hline \hline
\multicolumn{3}{|c|}{\textbf{MINIMIZATION}}
\\ \hline 
(P3) &$\underset{\B{s}}{\text{min}}$ &$\EX{\|\B{s}\|^2} $ 
\\
{} &s.t. &$\Delta(\B{s},\B{h}_B,\B{h}_E)\geq$ a threshold
\\ \hline 
(P4) &$\underset{\B{s}}{\text{min}}$ &$C_{\textrm{E}}(\B{s}, \B{h}_E)$
\\
{} &s.t. &$C_{\textrm{B}}(\B{s},\B{h}_B)\geq$ a threshold
\\
{} &{} &$\EX{\|\B{s}\|^2} \leq P_{max}$
\\ \hline 
\end{tabularx}
\label{tab: opt probs}
\end{table}

Due to the random nature of channels, the \emph{instantaneous} quantities $C_B(\B{s},\B{h}_B)$, $C_E(\B{s},\B{h}_E)$ and $\Delta(\B{s},\B{h}_B,\B{h}_E)$ will always {fluctuate} with $\B{h}_B$ and $\B{h}_E$. As such, it is practically {more promising} to deal with their expectations. Let $C_{B,avr}(\B{s})$, $C_{E,avr}(\B{s})$ and $\Delta_{avr}(\B{s})$ {represent} the expected value functions of $C_B(\B{s},\B{h}_B)$, $C_E(\B{s},\B{h}_E)$ and $\Delta(\B{s},\B{h}_B,\B{h}_E)$ over $\B{h}_B$ and $\B{h}_E$, we have
\begin{align}
C_{B,avr}(\B{s}) &= \EXs{\B{h}_B}{ C_B(\B{s},\B{h}_B) } ,
\\
C_{E,avr}(\B{s}) &= \EXs{\B{h}_E}{ C_E(\B{s},\B{h}_E) } ,
\\
\Delta_{avr}(\B{s}) &= \EXs{\B{h}_B,\B{h}_E}{ \Delta(\B{s},\B{h}_B,\B{h}_E) } ,
\end{align}
where $\EXs{\B{z}}{func(\B{z})}$ denotes the expectation operator that calculates the expected value of a certain function $func(\B{z})$ over a certain random vector $\B{z}$. Note that $C_{B,avr}(\B{s})$, $C_{E,avr}(\B{s})$ and $\Delta_{avr}(\B{s})$ are still functions of $\B{s}$. 

\begin{table}[!t]
	\centering 
	\caption{Types of Stochastic Optimization Problems in PHY security}
	\renewcommand{\arraystretch}{1.5}
	\begin{tabularx}{.7\linewidth}{|ccX|}
		\hline 
		\multicolumn{3}{|c|}{\textbf{MAXIMIZATION}}
		\\ \hline 
		(Q1) &$\underset{\B{s}}{\text{max}}$ &$\Delta_{avr}(\B{s})$ 
		\\
		{} &s.t. &$\EX{\|\B{s}\|^2} \leq P_{max}$ 
		\\ \hline
		(Q2) &$\underset{\B{s}}{\text{max}}$ &$C_{B,avr}(\B{s})$ 
		\\
		{} &s.t. &$C_{E,avr}(\B{s})\leq$ a threshold 
		\\
		{} &{} &$\EX{\|\B{s}\|^2} \leq P_{max}$ 
		\\ \hline \hline \hline
		\multicolumn{3}{|c|}{\textbf{MINIMIZATION}}
		\\ \hline
		(Q3) &$\underset{\B{s}}{\text{min}}$ &$\EX{\|\B{s}\|^2} $ 
		\\
		{} &s.t. &$\Delta_{avr}(\B{s})\geq$ a threshold 
		\\ \hline
		(Q4) &$\underset{\B{s}}{\text{min}}$ &$C_{E,avr}(\B{s})$ 
		\\
		{} &s.t. &$C_{B,avr}(\B{s})\geq$ a threshold
		\\
		{} &{} &$\EX{\|\B{s}\|^2} \leq P_{max}$
		\\ \hline 
	\end{tabularx}
	\label{tab: opt probs: modified}
	\vspace{0.05cm}
\end{table}
If the distributions of $\B{h}_B$ and $\B{h}_E$ are known, then the expected values $C_{B,avr}(\B{s})$, $C_{E,avr}(\B{s})$ and $\Delta_{avr}(\B{s})$ may be derived by integrating {- i.e. averaging -} the instantaneous quantities $C_B(\B{s},\B{h}_B)$, $C_E(\B{s},\B{h}_E)$ over the domains of $\B{h}_B$ and $\B{h}_E$.
However, in practice, the three distributions of the channels $\B{h}_B$ and $\B{h}_E$ may not be {known}. Instead of using integration, the expected values may be estimated based on measurement. Let $\B{h}_B^{[t]}$ and $\B{h}_E^{[t]}$ be the {measured} value of $\B{h}_B$ and that of $\B{h}_E$ at the $t$-th time slot. Note that $\B{h}_B^{[t]}$ and $\B{h}_E^{[t]}$ are empirical observations, and hence they are given. After $T$ time slots, $\Delta_{avr}(\B{s})$, $C_{B,avr}(\B{s})$ and $C_{E,avr}(\B{s})$ can be approximated by the \emph{sample means} (or \emph{sample averages}) as follows:
\begin{align}
C_{B,avr}(\B{s}) &= \frac{1}{T}  \sum_{t=1}^{T} C_B(\B{s},\B{h}_B^{[t]}) ,
\\
C_{E,avr}(\B{s}) &= \frac{1}{T} \sum_{t=1}^{T} C_E(\B{s},\B{h}_E^{[t]}) ,
\\
\Delta_{avr}(\B{s}) &= \frac{1}{T} \sum_{t=1}^{T} \Delta(\B{s},\B{h}_B^{[t]},\B{h}_E^{[t]}) .
\end{align} 
From a practical point of view, the objective functions, as well as constraints, in the optimization problems (P1)-(P4) {of Table \ref{tab: opt probs}} should be modified. For example, they can be reformulated into new optimization problems (Q1)-(Q4), which are shown in Table \ref{tab: opt probs: modified}. 

In general, the problems (Q1)-(Q4) {of Table \ref{tab: opt probs: modified}} are non-convex and
computationally  challenging to solve. The formulation of (Q1)-(Q4) can be generalized and expressed in the form of a minimization problem as follows:\footnote{Since maximizing $f(\B{x})$ is equivalent to minimizing $(-1) f(\B{x})$, the maximization problems (Q1)-(Q2) can also be transformed into minimization problems.}
\begin{subequations}\label{Original OPT}
	\begin{alignat}{2}
	\underset{ \B{z} }{\text{min}} 
	\quad
	&\mathcal{F}(\B{z}) = \frac{1}{T} \sum_{t=1}^T f_t(\B{z}) 
	\label{ori opt: objective} \\
	\text{s.t.} \quad
	& \B{z} \in \mathcal{S} ,
	\label{ori opt: constraints} 
	\end{alignat}
\end{subequations} 
where $\B{z}$ is a \emph{complex-valued} random vector, $\mathcal{S}$ is some constraint domain in the \emph{complex} field, $\mathcal{F}(\B{z})$ is the overall objective function, $f_t(\B{z})$ is a certain function of $\B{z}$, and $T$ is still the number of {training} examples. Note that the non-convex problem \eqref{Original OPT} {is routinely} encountered in many technological fields. 

\begin{table*}[!t]
	\centering
	\caption{In-depth Research on Using Neural Networks for Solving Optimization Problems}
	\renewcommand{\arraystretch}{1.5}
	\begin{tabular}{|c|c|c|c|c|}
\hline \cline{1-5}
		\multirow{2}{*}{ \shortstack{\textbf{Stochastic}\\\textbf{Optimization}} }
		& \multirow{2}{*}{ \shortstack{\textbf{Constrained}\\\textbf{Optimization}} } 
		& \multirow{2}{*}{ \shortstack{ \textbf{Gradient Descent} \\ \textbf{(GD) Methods} } }
		& \multirow{2}{*}{ \textbf{Content to be Emphasized } }
		& \multirow{2}{*}{ \textbf{Papers} }
\\ 
{} 
& {}  
& {}  
& {} 
& {} 
\\ \hline \cline{1-5}
{ } 
& $\checkmark$ 
& Projection 
& Projection neural network 
&\cite{Projection-NN_Jin2019}    
\\ \hline   
{ }
& $\checkmark$ 
& Projection 
& Projection neural network  
&\cite{Projection-NN_Yang2007}    
\\ \hline 
{ }
& $\checkmark$ 
& Projection  
& Projection neural network  
&\cite{Projection-NN_Effati2007}    
\\ \hline 
{ }
& $\checkmark$ 
& Projection  
& RNN based on Projection Method  
&\cite{Feedback_NN-Constrained_OPT_Yang2008}    
\\ \hline 
{ }
& $\checkmark$ 
& Projection  
& Projection neural network  
&\cite{Projection-NN_Xia2002}    
\\ \hline 
{ }
& $\checkmark$ 
& Projection  
& Projection neural network  
&\cite{Projection-NN_Liu2015}    
\\ \hline 
{ }
& $\checkmark$ 
& \textit{Not to be mentioned} 
& Modified RNN based on Karush--Kuhn--Tucker Conditions  
&\cite{AlaeddinMalek2015} 
\\ \hline
{ }
& $\checkmark$ 
& Projection 
& Complex-valued Projection neural network  
&\cite{Complex_Projection_NN_Zhang2015}    
\\ \hline \cline{1-5}
$\checkmark$ 
& { }
& SGD 
& Showing the computational efficiency in large-scale ML problems 
&\cite{GD_NN_Bottou2010}    
\\ \hline
$\checkmark$ 
& { }
& ADAM 
& Proposing ADAM and showing its computational efficiency 
&\cite{ADAM_Kingma2015}    
\\ \hline
$\checkmark$ 
& { }
& YOGI 
& Proposing YOGI and proving it superior to ADAM 
&\cite{STO_SGD_NN_Zaheer2018}    
\\ \hline
$\checkmark$ 
& { }
& Online, Batch 
& Presenting online learning algorithms and analyzing their convergence 
&\cite{STO_NN_Bottou1998}  
\\ \hline \cline{1-5} 
$\checkmark$ 
& $\checkmark$ 
& AdaGrad 
& Proposing a family of update methods that use the geometry of data 
&\cite{AdaGrad_Duchi2011} 
\\ \hline
$\checkmark$ 
& $\checkmark$ 
& SGD + Projection  
& Proposing an algorithm for stochastic multiple objective optimization 
&\cite{Projected_STO_ML_Mahdavi2012}    
\\ \hline 
$\checkmark$ 
& $\checkmark$ 
& SGD + Projection  
& The effect of stochastic errors on distributed subgradient algorithms  
&\cite{Projected_STO_SundharRam2010}    
\\ \hline
$\checkmark$ 
& $\checkmark$ 
& SGD + Projection  
& Analyzing the optimality of SGD and improving its convergence rate 
&\cite{AlexanderRakhlin2012}    
\\ \hline
$\checkmark$ 
& $\checkmark$ 
& SGD + Projection  
& Proposing a variant of mini-batch SGD to enhance the convergence rate 
&\cite{Projection_OPT_NN_Li2014}      
\\ \hline \cline{1-5}
\end{tabular}
\label{table: STO OPT + NN + GD + papers}
\end{table*}

\subsection{NN-aided Optimization Solutions}\label{sub-sec: real-valued NNs}
To handle non-convex problems, a wide range of methods has been proposed. If a non-convex problem can be innerly  approximated by a  convex problem, the solution of the latter  is also a feasible point/sub-optimal solution   of the former. However, if a non-convex problem is relaxed into  a convex problem, the solution
of the latter  may not even be a feasible point  for the former.
In fact, there is no single universal method for efficiently solving all types of non-convex  problems.  {This statement is} true, when using neural networks for solving any optimization problem. {Explicitly,} specific network designs may be used for specific optimization problems. Table \ref{table: STO OPT + NN + GD + papers} lists several papers, which use neural networks {for solving} optimization problems. 

{In order to be able to harness} neural networks to solve our security problems, we may desire to transform (or relax) \eqref{Original OPT} into the following equivalent (or approximate) optimization problem:
\begin{subequations}\label{STO OPT}
	\begin{alignat}{2}
	\underset{ \B{w} }{\text{min}} 
	\quad
	& \mathcal{L}(\B{w}) = \frac{1}{T} \sum_{t=1}^T \ell_t(\B{w}) 
	\label{sto opt: objective} \\
	\text{s.t.} \quad
	& \B{w} \in\mathcal{W} ,
	\label{sto opt: constraints} 
	\end{alignat}
\end{subequations} 
where $\B{w}$ is a \emph{real-valued} vector containing the \emph{weights and biases} of a neural network, $\mathcal{W}$ is a closed convex set in the \emph{real} field, $\mathcal{L}(\B{w})$ is the overall loss function, $\ell_t(\B{w})$ is the loss function corresponding the $t$-th example, and $T$ is the training data size, i.e., the number of examples in the training data. The desire for $\B{w}$ {to be} a real-valued vector {arises} from the fact that most of the existing neural networks {have been} developed for solving optimization problems {associated} with real variables. Hence, there is a {paucity of} studies using neural networks for solving optimization problems {having} complex variables, {except for the work of Zhang} \textit{et al.} \cite{Complex_Projection_NN_Zhang2015}. 

As such, instead of directly dealing with an optimization problem w.r.t. the \emph{complex-valued} vector $\B{z}$, we {might prefer} using a neural network to solve another optimization problem w.r.t. the \emph{real-valued} parameter vector $\B{w}$. Once the optimal (or near-optimal) solution of $\B{w}$ has been found, $\B{z}$ in the original problem will also be updated accordingly. However, there may or may not be a functional relationship between $\B{w}$ and $\B{z}$. 

It is also a matter of \emph{utmost importance} that network designers {establish} the relationship between the complex-valued vector $\B{z}$ and the real-valued vector $\B{w}$, because this relationship will also show the relationship between the constraint $\B{z}\in\mathcal{S}$ in \eqref{ori opt: constraints} and the constraint $\B{w}\in\mathcal{W}$ in \eqref{sto opt: constraints}. Similarly, it is challenging, {but} important to {establish} the relationship between the objective function $\mathcal{F}(\B{z})$ in \eqref{ori opt: objective} and the loss function $\mathcal{L}(\B{w})$ in \eqref{sto opt: objective}. For example, $\B{z}$ can be chosen to be a real-valued output vector that is dependent on $\B{w}$ \cite{He2019-PLS-ML} or independent of $\B{w}$ \cite{Kang2018-Deep}.

Last but not least, \eqref{STO OPT} {represents} a type of stochastic optimization that has been studied {both} in machine learning research as well as {in} applied mathematics for at least a decade \cite{STO_Ermoliev1988, Projected_STO_ML_Agarwal2011, Projection_OPT_NN_Li2014, ADAM_Kingma2015,  Projected_STO_NN_Hardt2016}. 
As seen in Table \ref{table: STO OPT + NN + GD + papers}, stochastic optimization {problems may be} solved with the help of neural networks and gradient descent methods (see \cite{GD_NN_Bottou2010, STO_SGD_NN_Zaheer2018, ADAM_Kingma2015, Projected_STO_ML_Mahdavi2012, Projected_STO_SundharRam2010, AlexanderRakhlin2012, Projection_OPT_NN_Li2014, STO_NN_Bottou1998} for more details). 
{Given} the importance of gradient (descent) methods, the following discussions {will unveil the} the role of them in dealing with {both}: i) stochastic \emph{unconstrained} optimization and ii) stochastic \emph{constrained} optimization {problems}. 

\subsubsection{\textbf{Stochastic Optimization without Constraints}}
For {ease of exposition}, in this sub-section, we temporarily remove the constraint $\B{z}\in\mathcal{S}$ from the stochastic optimization problem \eqref{STO OPT}. With this in mind, we will discuss how neural networks can deal with stochastic optimization problems of the form
\begin{subequations}
\begin{alignat}{2}
\underset{ \B{w} }{\text{min}} 
\quad
& \mathcal{L}(\B{w}) = \frac{1}{T} \sum_{t=1}^T \ell_t(\B{w}) .
\label{opt without constraints} 
\end{alignat}
\end{subequations} 
According to \cite{STO_NN_Bottou1998, GD_TRICKS_Bottou2012, Projected_STO_ML_Mahdavi2012}, we can find the optimal (or near-optimal) solution of the parameter vector $\B{w}$ by {relying on the classical} gradient descent. 
In connection with each gradient descent method, the update of $\B{w}$ at the $(i+1)$-st iteration, denoted by $\B{w}_{i+1}$, follows a different rule, as shown below:
\begin{itemize}
	\item Gradient descent (GD) \cite{STO_NN_Bottou1998, GD_TRICKS_Bottou2012}: 
		\begin{align}
			\B{w}_{i+1} = \B{w}_i - \gamma_i \frac{1}{T} \sum_{t=1}^{T} \nabla_{\B{w}} \ell_t(\B{w}_i) ,
		\end{align} 
	where $\gamma_i > 0$ is the learning rate.
	\item Stochastic gradient descent (SGD) \cite{STO_NN_Bottou1998, GD_TRICKS_Bottou2012, Projected_STO_ML_Mahdavi2012, STO_SGD_NN_Zaheer2018}:
		\begin{align}
			\B{w}_{i+1} = \B{w}_i - \gamma_i  \nabla_{\B{w}} \ell_i(\B{w}_i) .
		\end{align}
	\item Mini-batch gradient descent \cite{Mini-batch_Masters2018}:
		\begin{align}
			\B{w}_{i+1} = \B{w}_i - \gamma_i \frac{1}{\tau} \sum_{t=1}^{\tau} \nabla_{\B{w}} \ell_t(\B{w}_i) ,
		\end{align}
		where $\tau\leq T$ is the number of training examples in a small subset of the full training dataset. In the case of the mini-batch GD, the training dataset of $T$ samples is divided into many subsets, and each update corresponds to a subset. Setting $\tau = 1$ leads to the case of SGD, while setting $\tau=T$ leads to the gradient descent {scenario}. 
\end{itemize}
{Apart from} the aforementioned gradient descent methods, there are also other gradient-based methods, such as the second-order gradient descent \cite[eq. (3)]{GD_NN_Bottou2010}, the second-order SGD \cite[eq. (5)]{GD_NN_Bottou2010}, ADAM \cite{ADAM_Kingma2015}, and YOGI \cite{STO_SGD_NN_Zaheer2018}, just to name a few. 

\begin{figure}[t!]	\centerline{\includegraphics[width=0.95\linewidth]{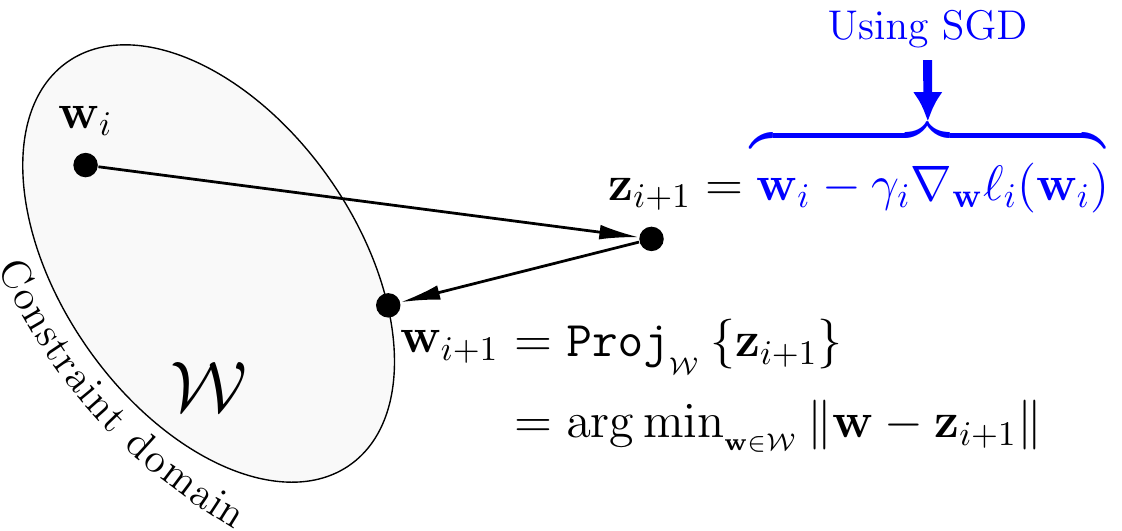}}
	\caption{An illustration of projection gradient descent. The projection-based update rule includes two steps as follows: Starting at a point $\mathbf{w}_i \in \mathcal{W}$, we use a gradient method (e.g., SGD) to find some intermediate point $\B{z}_{i+1}$. Then, we project $\B{z}_{i+1}$ onto the domain $\mathcal{W}$ to find the updated point $\mathbf{w}_{i+1}$ that satisfies the constraint $\mathbf{w}_{i+1} \in \mathcal{W}$.}
	\label{fig: Projection GD}
\end{figure}

\subsubsection{\textbf{Stochastic Optimization under Constraints}}
A specific type of feedback/recurrent neural networks (RRNs), termed as \emph{projection neural networks}, have been {popularly} used for solving constrained optimization over many years \cite{Projection-NN_Xia2002, Projection_NN_OPT_Hu2006, Projection-NN_Jin2019}. As for stochastic constrained optimization problems (e.g., \eqref{STO OPT}), projection neural networks can also be used {for finding} optimal (or near-optimal) solutions. In principle, the constraint $\B{w} \in\mathcal{W}$ in \eqref{STO OPT} can be handled by the projection method {of} \cite{Projected_STO_ML_Mahdavi2012, Projected_STO_SundharRam2010, AlexanderRakhlin2012, Projection_OPT_NN_Li2014}. 

Figure \ref{fig: Projection GD} illustrates the update rule of a projection neural network. To be more particular, the update rule includes two steps: 
\begin{itemize}
	\item Let us assume that we {commence from} a point $\B{w}_i$ at the $i$-th iteration. In the first step, a gradient descent method is used {for finding} an intermediate point, namely $\B{z}_{i+1}$. Note that the point $\B{z}_{i+1}$ does not necessarily fall within the domain $\mathcal{W}$ (i.e., it is possible to have $\B{z}_{i+1}\notin\mathcal{W}$). 
	\item In the second step, the projection method is used {for projecting} the intermediate point $\B{z}_{i+1}$ onto the domain $\mathcal{W}$. Let $\B{w}_{i+1}$ be the projected point. It is {plausible} that $\B{w}_{i+1}$ is also the updated point at the $(i+1)$-st iteration, because the constraint $\B{w}_{i+1}\in\mathcal{W}$ is satisfied.
\end{itemize}
This 2-step update process is also known as the projected gradient descent (PGD) {technique} \cite{soltanolkotabi2017learning, andrychowicz2016learning}. As an example, let us assume that in the first step, we use the SGD {for updating} the intermediate point, i.e,
\begin{align}
	\B{z}_{i+1} = \B{w}_i - \gamma_i \nabla_{\B{w}} \ell_i (\B{w}_i).
\end{align}
Then we use the projection method {for updating} the parameter vector as follows:
\begin{align}
	\B{w}_{i+1} &= \texttt{Proj}_{ {}_{\mathcal{W}} } \left\{ \B{z}_{i+1} \right\}
	\nonumber \\
	&= \arg\min_{\B{w}\in\mathcal{W}} \| \B{w} - \B{z}_{i+1} \| 
\end{align}
where $\texttt{Proj}_{ {}_{\mathcal{W}} } \{ \cdot \}$ denotes the operator {carrying out the} projection onto $\mathcal{W}$.

\subsection{Recent Advances and Future Directions}\label{sub-sec: NNs: complex, Pareto}
\subsubsection{Security Optimization by Neural Networks}
{The} variables {encountered} in optimization problems may be real-valued or complex-valued. Hence, we will classify the optimization problems {considered} into these two types: 
\begin{itemize}
	\item \emph{Real}-variable optimization (real-OPT) problems, whose variables belong to the real field;
	\item \emph{Complex}-variable optimization (complex-OPT) problems, whose variables belong to the complex field. 
\end{itemize}
{Indeed, the employment of} neural networks for solving optimization problems has been an active {field} for years. However, most of the optimization problems {solved have been} \emph{real}-valued problems \cite{Lee2019-applied-NNs-for-OPT, Eisen2019-applied-NNs-for-OPT}. {Only a few authors} consider the use of neural networks for solving \emph{complex}-valued optimization problems \cite{Complex_ProjectionNN_Savitha2013, Complex_Projection_NN_Zhang2015}. 

{Similarly to the} optimization problems, we can also divide {the family of} neural networks into {a pair of} categories:
\begin{itemize}
	\item \emph{Real}-valued neural networks (namely, real-NNs);
	\item \emph{Complex}-valued neural networks (namely, complex-NNs).
\end{itemize}
Naturally, the first category relates to all neural networks designed for solving real-OPT problems. {By contrast}, the second category {encompasses the} neural networks designed for solving complex-OPT problems. As a matter of fact, {in many fields of application,} real-OPT problems are more likely to be {encountered, but} in the field of communications, {typically the opposite is true. Hence, there is a pressing need for further research on solving the unsolved open optimization problems (including real-OPT and complex-OPT problems) of} the communications comminity. {To elaborate a little further}, {some authors have applied} real-NNs for solving real-OPT problems to improve the performance of communication systems \cite{Kang2018-DeepCNN, He2019-PLS-ML}, {but there is a paucity of literature on the employment} of complex-NNs for solving complex-OPT problems in communication systems. 

\begin{figure*}[t!]	\centerline{\includegraphics[width=.9\linewidth]{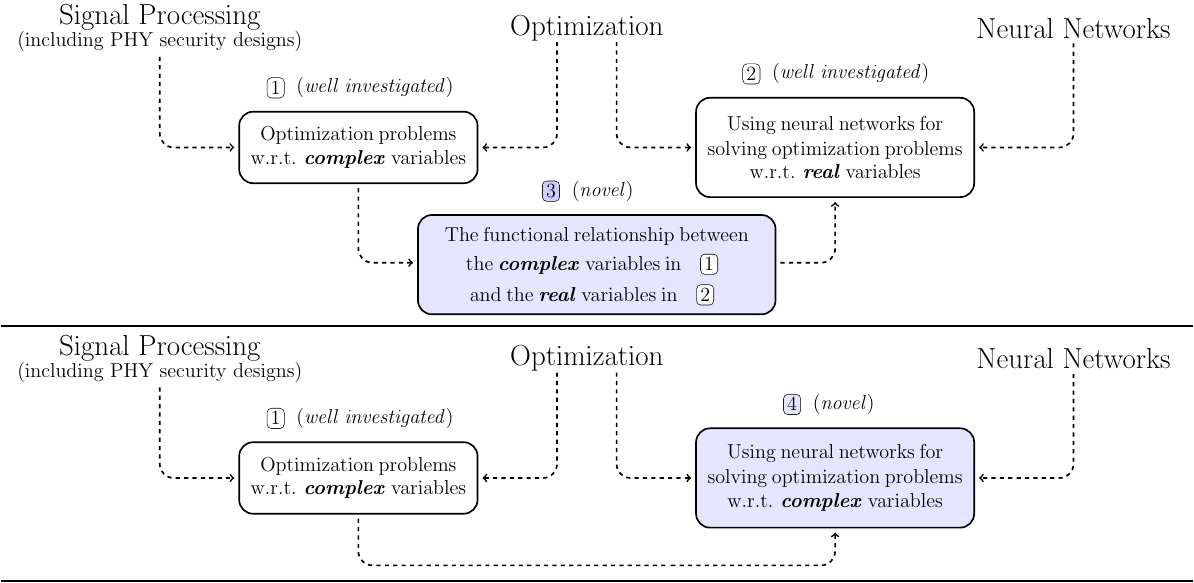}}
	\caption{In communications, we are most likely to deal with complex variables rather than real variables in optimization problems. Thus, future works are envisaged to happen with $2$ possibilities: (i) find a way to transform the original optimization problems w.t.r complex variables into new ones that conventional neural networks can solve, and (ii) use new types of neural networks that can deal with complex numbers in order to solve the original optimization problems directly.}
	\label{fig: open prob: opt + NN}
\end{figure*}

\begin{itemize}
	\item Using real-NNs for solving complex-OPT problems {requires further research for transforming complex-OPT problems into real-valued ones, so that real-NN may be harnessed for solving them.} The upper part of Figure \ref{fig: open prob: opt + NN} illustrates this issue. 
	\item {However, it is} more natural to use complex-NNs for directly solving complex-OPT problems \cite{Complex_Projection_NN_Zhang2015}. {Having said that, substantial future research is required on} complex-NNs \cite{Complex_ProjectionNN_Savitha2013, Complex_Projection_NN_Zhang2015} {before they can be harnessed} for solving complex-OPT problems {routinely encountered by} the communications community. The lower part of Figure \ref{fig: open prob: opt + NN} illustrates this issue. 
\end{itemize}

Since communication system designs include PHY security, it is anticipated that real-NNs, as well as complex-NNs will help enhance their {security in the face of uncertainty, when the ability to learn from and adapt to the environment is of crucial importance.}

\subsubsection{Security Optimization {for satisfying} Multiple Objectives}
When it comes to the formulation of optimization for a system, it may be {desirable} to {simultaneously} optimize multiple objectives even when they are conflicting.  Multiple-objective optimization can be {formulated} as follows:
\begin{subequations}\label{MOO}
	\begin{alignat}{2}
	\underset{ \B{z} }{\text{min}} 
	\quad
	& \left[ f_1^{obj}(\B{z}), f_2^{obj}(\B{z}), \ldots, f_{M_{obj}}^{obj}(\B{z}) \right]
	\label{MOO: objective} \\
	\text{s.t.} \quad
	& \B{z} \in \mathcal{S} ,
	\label{MOO: constraints} 
	\end{alignat}
\end{subequations} 
where $f_m^{obj}(\cdot)$ (with $m\in \{1,2,\ldots,M_{obj}\}$) is an objective function in the set of all objectives, while $\mathcal{S}$ is some constraint domain in the complex field.
{To elaborate,} multiple-objective optimization (MOO) is different from the single-objective optimization problems presented in Tables \ref{tab: opt probs}--\ref{tab: opt probs: modified} of Section \ref{SEC: Design - NN}. 
{Explicitly,} the single-objective optimization problems of Section \ref{SEC: Design - NN} {simplify the complex multi-objective real-life problems by simply using the} conflicting objectives {as} constraints. 
By contrast, the goal of multiple-objective optimization is to \textit{jointly} optimize several objectives at the same time, where each objective $f_m^{obj}(\cdot)$, $m\in \{1,2,\ldots,M_{obj}\}$ {represents a different} performance metric. {Naturally, the solution-space or search-space of this problem continues to grow upon including more metrics, which eventually renders the problem intractable.}

{Hence}, multiple-objective optimization {problems are challenging} to deal with and the optimal solution {may not even} exist.
Having said that, it is possible to {circumvent the challenges of} multiple-objective optimization using the Pareto optimality concept \cite{Pareto_Fei2017} {relying on} a set of solutions {which constitute} the \emph{Pareto front} {of all optimal solutions. For example, we} can improve a performance metric (e.g., the capacity of a legitimate channel), but another performance metric will {be degraded} (e.g., the capacity of a wiretap channel will increase). {Broadly speaking,} the Pareto {front is the collection of all optimal solutions, where none of the metrics may be improved without degrading at least one of the others.}

To find the Pareto front, typically \emph{bio-inspired metaheuristic} algorithms are employed \cite{Pareto_Fei2017, Pareto_Bio-inspired_Mirjalili2017, Pareto_Bio-inspired_Kar2016}. Acording to \cite{Pareto_Fei2017}, common bio-inspired metaheuristic algorithms include the following main categories: \emph{evolutionary algorithms}, \emph{swarm intelligence algorithms}, \emph{neural networks}, \emph{reinforcement learning}, \emph{fuzzy logic}, just to name a few. For example, the combination of neural networks and the Pareto approach is studied in \cite{Pareto_ML_Hussien2003}, while the combination of reinforcement learning and the Pareto approach is presented in \cite{Pareto_ML_Tozer2017}. {Since} machine learning {itself may be viewed as} a multiple-objective {optimization tool, its} pairing {with} the Pareto approach is a natural {marriage} \cite{Pareto_ML_Jin2008}. 
  
Multiple-objective optimization {is eminently suitable for} network design \cite{Pareto_ML_GA_Chang2019}, but it is rarely used in PHY security designs. {Naturally,} in PHY security, at least one of the objectives in the optimization problem \eqref{MOO} must be related to a security metric \cite{Pareto_Security_Wu2017}. 
{Indeed, wireless} networks pose many challenges related to the maximization of the data rate, the minimization of consumed power, the maximization of security level, the minimization of latency and so on. These goals are hard to achieve at the same time, because they are often conflicting. 
{For example, the bit error rate of a wireless system may be reduced by increasing the transmit power, but this increases the eavesdropping probability, which results in a security vs. information trade-off} \cite{Security-reliability_Zou1}. The trade-off between security and delay {was considered in} \cite{Security-reliability_tradeoff}. 
	
	


\begin{figure*}[t!]	\centerline{\includegraphics[width=.95\linewidth]{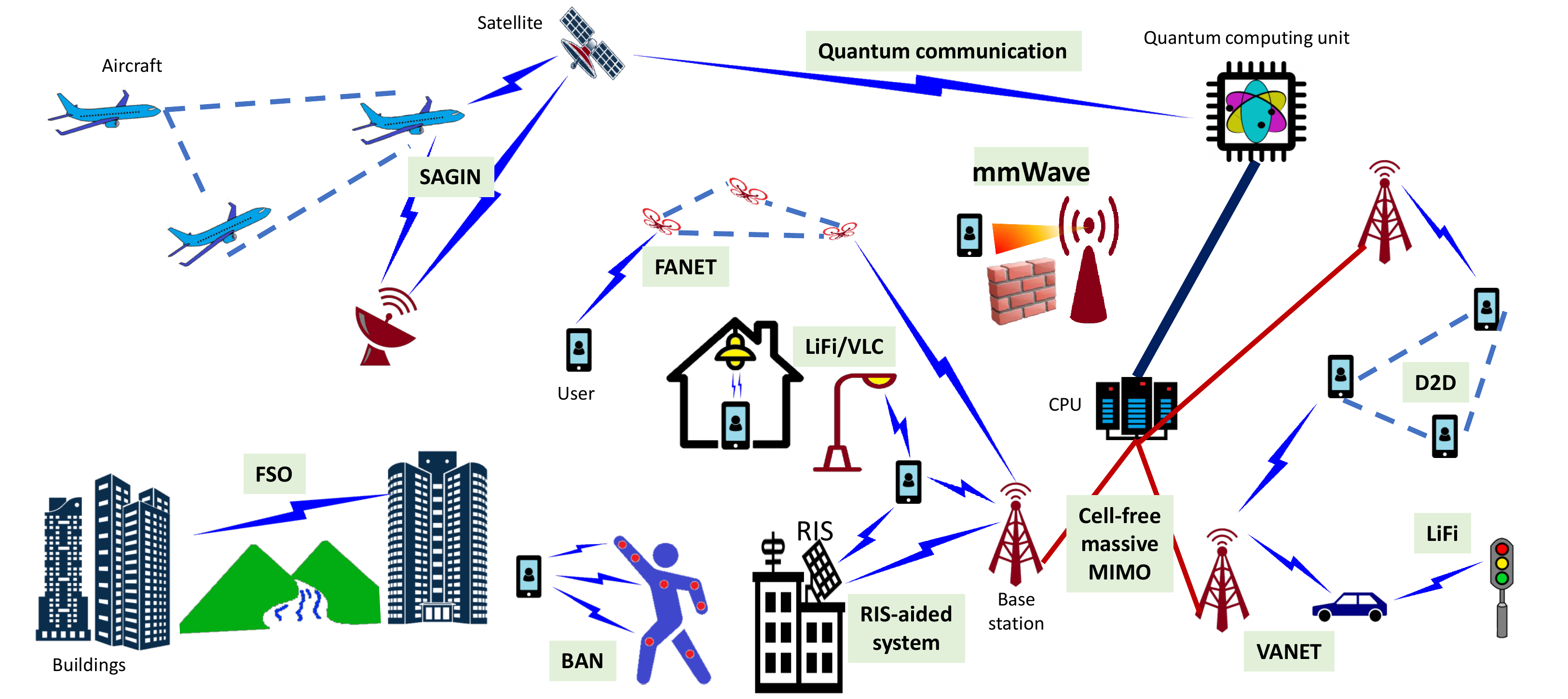}}
	\caption{A visualization of future networks.}
	\label{fig: future systems}
\end{figure*}

\section{Further Discussion: ML-aided PHY Security for Future Communication Systems}\label{sec: future}
When new wireless networks are rolled out, their security level {has to be} improved. Given that ML techniques {are gradually finding their way into} the new generation of wireless networks, the investigation of ML-aided PHY security in those networks continues to be important part of future research. Below {we critically appraise a number of} emerging solutions in wireless networks:
\begin{itemize}
	\item \textbf{Reconfigurable intelligent surfaces (RISs)}: It is expected that RISs will be deployed at locations near the {base stations for mitigating the LoS blockages} \cite{RIS-survey-2021}. {They} are capable of creating additional {reflected} propagation paths and {beneficially} controlling the phase shifts of reflected signals, thus making robust connections between a transmitter and a receiver. In terms of security, it {is still} an open question {how to orchestrate} the coordination of RISs and ML techniques for improving the security level of wireless systems, {although some insights} have been {provided in} \cite{RIS-ML-Security-Guo2021, RIS-ML-Security-Jiang2022, RIS-ML-Security-Tiep2022}. However, there is a {pressing} need to perform more extensive investigations on the security performance of ML-aided RISs.
	\item \textbf{Millimeter wave (mmWave)}: While most existing communication systems use carrier frequencies below 6 GHz, millimeter wave (mmWave) {systems} employ a wide range of frequency spectrum {spanning} from 30 GHz to 300 GHz \cite{mmWave_survey_Robert2016, mmWave_survey_Rappaport2015}. {Their} PHY security has been {characterized in} \cite{mmWave_security_Wang2016, mmWave_security_Vuppala2018}, which ML has been {harnessed in} \cite{mmWave_ML_Huang2019, mmWave_ML_Khan2019}. However,
	there is {a paucity of literature on} ML-aided PHY security in mmWave {systems, apart from \cite{mmWave_security_ML_Wang2020}, \cite{ mmWave_security_ML_Xiao2018}. Briefly,} logistic regression is proposed in \cite{mmWave_security_ML_Wang2020} to detect eavesdropping attacks in the uplink, while reinforcement learning is used in \cite{ mmWave_security_ML_Xiao2018} to deal with jamming. 
	\item \textbf{Visible light communication (VLC)}: VLC technology {modulates} visible light {emanating} from light-emitting diodes (LEDs) \cite{VLC_Survey_LED_Indoor} {used} for illumination at low cost.{Since} the light from LEDs does not {propagate through walls}, VLC is robust to interference. Thus, VLC is a potential candidate for future indoor systems. {However,} due to the broadcast nature of VLC in the downlink, it is vulnerable to eavesdroppers. {Hence}, PHY security has also been investigated in the context of VLC \cite{VLC_Security_Mostafa2015, VLC_Security_SSK_Wang2018, VLC_Security_Wang2018}. Having said that, a full investigation of {ML-aided} PHY security in VLC is still largely open.   
	\item \textbf{Light fidelity (LiFi)}: 
	While VLC only uses the visible light spectrum, LiFi {additionally} exploits the infrared and the ultraviolet {bands} \cite{VLC_LiFi_Haas2018, Optical_Wireless_Commun_2018}. {However, the} ML-aided PHY security of LiFi {is an open area.} 
	\item \textbf{Cell-free massive MIMO}: As a variant of distributed massive MIMO {systems}, cell-free massive MIMO {has numerous benefits, especially} in terms of throughput {fairness} \cite{Cell-free_AHien}, {despite its low complexity.} Its PHY security is investigated in
	\cite{Cell-free_Hoang2018}. {As a further advance,} \emph{federated learning} is proposed for cell-free massive MIMO in \cite{vu2019cellfree}, but the investigation of ML-aided PHY security is still {in its infancy}. 
	\item \textbf{Body area networks}: A wireless body area network is typically comprised of medical sensors that are placed on the body to measure physiological signals. A practical wireless body area network is expected to collect data from a large pool of patients and use ML algorithms for analyzing the health and needs of patients. {Indeed,} personalized healthcare through mobile wearable devices is expected to revolutionize the future of healthcare. {Furthermore, the authors of} \cite{BAN_ML_Mohamed2018} {rely on the} RSSI as the feature of data and compare different ML algorithms (i.e., decision tree, SVM, k-NN and neural network) {in the context} of gait authentication. {However}, the large-scale deployment of wireless body area networks must meet stringent security policies and requirements. Once a wireless body area network is entitled to use confidential data for health surveillance {relying on} statutory medical services, PHY security {becomes a pivotal} issue \cite{BAN_Security_GameTheory_Moosavi2016, BAN_Security_CompressedSensing_Dautov2016}. Thus, the design of body area networks has to ensure that the user database is not leaked to {illegitimate} users or organizations. {Hence}, ML-aided PHY security {constitutes} an exciting domain of research.
	\item \textbf{Space-air-ground communication}: {This is a topical research area} \cite{Satellite_Reinforcement_survey_Ferreira2019, UAV_survey_Li2019}, {but there is limited literature on} simultaneously considering all three network segments \cite{SAGIN_Survey_Huang2019, SAGIN_Survey_Liu2018}. 
	\item \textbf{Quantum communications}: 
	Quantum-aided communication systems have {rapidly evolved in recent years} \cite{Quant_Survey_Hosseinidehaj2019, Quant_Satellite_commun_1, CB-QSDC_classical_XOR, Quant_Air-to-Ground_Nauerth2013}.  
	In terms of security, quantum cryptography {relies on} a range of secure protocols such as, quantum key distribution \cite{Quant_Satellite_commun_1}, quantum secret
	sharing, quantum secure direct communication \cite{ Experimental_QSDC_single_photons}, and controlled bidirectional quantum secure direct communication \cite{CB-QSDC_classical_XOR}. These quantum cryptography protocols {have also found practical applications} \cite{Experimental_QSDC_single_photons, Quant_Proof_Gottesman2003}. 
	In terms of ML, a number of quantum-based learning algorithms have been developed to deal with critical problems in learning from data \cite{Quant_ML_Schuld2019, Quant_ML_Schuld2015}. {The family of popular} quantum machine learning algorithms includes quantum k-NN \cite{mezquita2020review}, quantum SVMs \cite{rebentrost2014quantum}, quantum neural networks \cite{schuld2014quest}, quantum decision tree \cite{lu2014quantum-DT}, just to name a few. Although quantum machine learning is still in the early stage of development, quantum machine learning algorithms {are capable of} speeding up computational processes, especially in learning from quantum-domain data  \cite{Quant_ML_Blank2019, Quant_ML_Afham2020, Quant_ML_Biamonte2017}. 
\end{itemize}
Other future systems include machine-type communications \cite{Machine-type-commun_Security}, free space optical communication \cite{FSO_Security}, and non-orthogonal multiple access systems \cite{NOMA_VLC_Security}, {but} ML-aided PHY security still remains a largely open domain of research. This is because the {benefit} of ML in PHY security has not been fully {documented and} the emergence of new communication systems will continue to widen the avenue for the investigation of ML-aided PHY security.

\section{Summary and Design Guidelines}\label{sec: conclusion}

\subsection{Summary}
In this work, we have summarized a variety of ML algorithms that can be employed in the context of PHY security. These range from typical supervised learning algorithms (i.e., $k$-NN, SVM, Random Forest, LDA) to typical unsupervised learning algorithms (i.e., $k$-means, OC-SVM, Isolation Forest, Hierarchical clustering). Neural networks, which can be classified as either supervised or unsupervised learning, has been also summarized. In parallel, we have discussed the state of the art in ML-aided PHY security by separately considering two aspects: ML-aided PHY authenticating and ML-aided PHY security design. Throughout the paper, we have shown that an ML classification algorithm can play two roles: i) it can classify the data for the authentication purpose, and ii) it can also take part in the process of selecting system components for secure transmission design. When it comes to security optimization, we have paid our special attention to the potential use of neural networks for solving optimization problems that are likely to be faced in designing PHY security strategies. Accordingly, we have shown and bridged the gap between complex-variable optimization in PHY security design and real-valued/complex-valued neural networks. Finally, we have presented the role of ML-aided PHY security in future communication systems. 

\subsection{Design Guidelines}
Regardless, whether we embark on the design of an authentication solution or a secure transmission strategy, the input data should be considered as the first step. More explicitly, the wireless data should be transformed into the input data for ML to interpret them. For example, an ML algorithm may be unable to directly use the modulated signals, because they are complex-valued; instead, it is necessary to decompose the complex-valued modulated signals into their amplitude and phase based representation in order for the ML algorithm to interpret and process them.

Given the input data, we will now briefly touch upon three representative authentication/security designs:

\begin{itemize}
    \item \textbf{Authentication:} 
      ML-based classification algorithms constitute prominent candidates for detecting security attacks, as well as for employment in authenticating devices. However, some ML algorithms work well on single-label data, while others only work on double-label data. Thus, it is necessary to determine the specific type of ML algorithm based on the nature of the available training datasets. A training dataset might contain attack-related samples if an attacker launched jamming/spoofing attacks in the past. Those attack-related samples may be labelled as ``attack'' if we have some information about the attacker, such as the CSI. On the other hand, without the CSI of the attacker, the attack-related samples in the dataset of interest might be treated as normal samples, which result in the wrong training dataset. Thus, based on the knowledge of the attacker, choosing carefully selected suitable data samples will allow us to choose a suitable ML algorithm for training. In this sense, given the attacker's CSI, binary/multi-label ML classification techniques may be used. On the other hand, without the attacker's CSI knowledge, both single-label ML classification techniques as well as anomaly detection techniques are more suitable candidates.
      
    \item \textbf{Component Selection:} 
      The family of ML-aided classification algorithms is also eminently suitable for optimizing antenna selection, relay selection and RIS selection. Let us consider an RIS-aided system as our example. When RISs are deployed in the field, they create additional propagation paths spanning from the source to the desired user for mitigating the effects of lin-of-sight blocking. However, in the presence of an eavesdropper, there will also be additional paths from the source to the eavesdropper, thus making the system vulnerable to malicious eavesdropping. Increasing the transmit power improves the transmission integrity between the source and the desired user, but also increases the eavesdropping probability. It is desirable to design the reflection coefficients of the RIS for ensuring that the benefits obtained by the legitimate user are higher than those obtained by the eavesdropper. By formulating an objective function based on the secrecy rate, bespoke ML classification algorithms can be trained to find the best set of RIS coefficients for maximizing the security rate.
      
    \item \textbf{Security Optimization:}
      While a real-valued neural network is routinely trained for minimizing its loss function $\mathcal{L}$, in wireless security design problems we often have to maximize the objective function, say $\Delta$. By assigning $\mathcal{L} = -\Delta$, a real-NN can be employed for dealing with security optimization. However, it is of vital importance to define the relationship between the outputs of the real-NN and the variables of the proposed security optimization problem. For example, if $z=z_1 + j\times z_2$ is a complex-valued variable to be optimized, then $z_1$ and $z_2$ may be represented by two outputs of the NN, since a real-NN can only process real-valued variables. Following the training process, the loss function $\mathcal{L}$ is minimized (or the objective function $\Delta$ is maximized), and the outputs $z_1$ and $z_2$ of the real-NN can be arranged into $z=z_1 + j\times z_2$ to find a near-optimal solution of the security optimization problem considered.

\end{itemize}

\bibliographystyle{IEEEtran}
\bibliography{IEEEabrv, Refs_DL, Refs_ML, Refs_New}

\end{document}